\documentclass[a4paper,11pt]{article}
\pdfoutput=1 

\usepackage{jheppub} 

\usepackage[T1]{fontenc} 
\usepackage{amsmath, amsthm, mathtools}
\usepackage{hyperref}
\usepackage{physics}
\def\R{\mathbb{R}}
\def\N{\mathbb{N}}
\def\C{\mathbb{C}}

\usepackage{caption}
\usepackage{subcaption}

\makeatletter
\newcommand{\vast}{\bBigg@{4}}
\newcommand{\Vast}{\bBigg@{5}}
\makeatother

\title{\boldmath Non-perturbative asymptotics of the eigenvalues of the spheroidal equation}

\author{M. Meynig}
\affiliation{Department of Physics, University of Connecticut,\\196A Auditorium Road, 06269, Storrs CT, United States of America}

\emailAdd{max.meynig@uconn.edu}

\abstract{
    New non-perturbative results on the eigenvalues of the spheroidal equation are presented.
    The results, found using an all orders WKB analysis, include a perturbative/non-perturbative (P/NP) relation as well as the first exponential correction to the perturbative series which is valid in certain regions of parameters.
    The quantum periods are also computed.
    }


\begin{document} 
\maketitle
\flushbottom

\section{Introduction}
\label{sec::Intro}

This article addresses the spectrum of the spheroidal equation
\begin{equation}\label{eq::SchrodingerSpheroidalEq}
    \left(- \frac{\hbar^2}{2} \derivative[2]{}{x} + \cos^2 x + \mu \csc^2 x - E \right) \psi(x) = 0.
\end{equation}
The new contributions presented here are a perturbative/non-perturbative (P/NP) relation, which shows that the non-perturbative terms of the transseries expansion for the eigenvalues of the spheroidal equation are entirely expressible in terms of perturbative quantities. 
Additionally, we provide systems of Picard-Fuchs differential operators which encode the quantum actions and thereby lay the foundations for analytic computations of the spectra of the spheroidal equation in multiple regions of parameter space and in various contexts.

\begin{figure}
    \centering
    \begin{subfigure}[a]{0.33\textwidth}
        \centering
        \includegraphics[width=\textwidth]{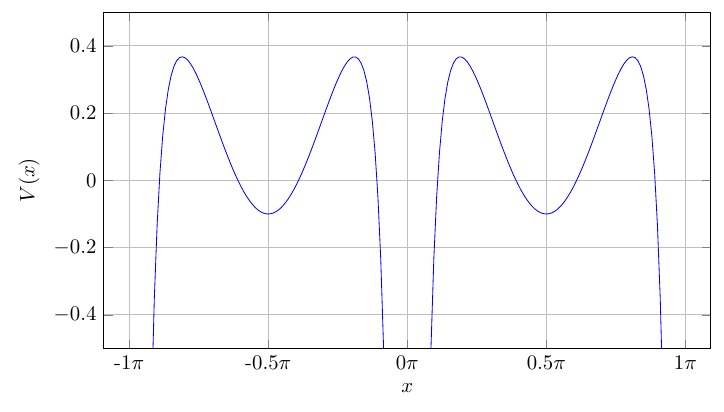}
        \caption{The region $-1<\mu<0$.}
    \end{subfigure}%
    \hfill
    \begin{subfigure}[a]{0.33\textwidth}
        \centering
        \includegraphics[width=\textwidth]{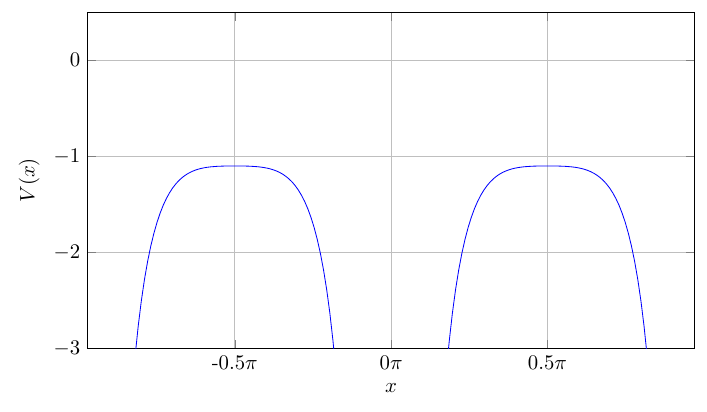}
        \caption{The region $\mu<-1$.}
    \end{subfigure}%
    \hfill
    \begin{subfigure}[a]{0.33\textwidth}
        \centering
        \includegraphics[width=\textwidth]{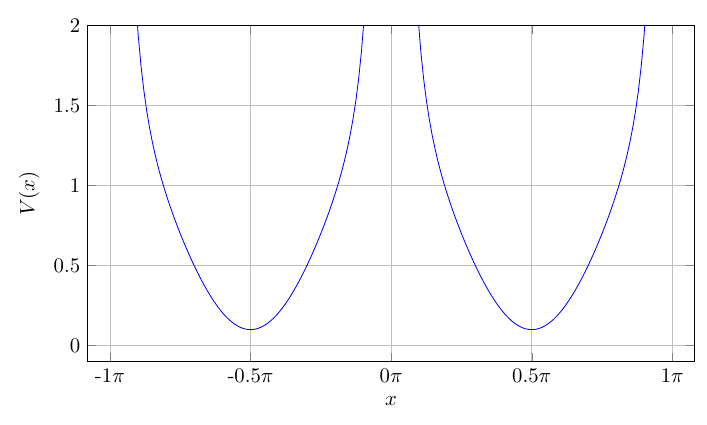}
        \caption{The region $0<\mu$.}
    \end{subfigure}%
    \caption{The spheroidal potential, $V(x) = \cos^2 x + \mu \csc^2 x$, plotted in the three distinct regions of the parameter $\mu$.
    This paper provides perturbative and non-perturbative expansions of the eigenvalue in the case $-1<\mu<0$. 
    \label{fig::PotentialRegions}}
\end{figure}

We explicitly compute a low energy expansion of the eigenvalue as well as the first non-perturbative correction in the region $-1<\mu<0$ (see figure \ref{fig::PotentialRegions}).
The non-perturbative terms provided are then checked using a numerical resurgent analysis.
Because of the dependence of the perturbative spectra on the level number and the parameter $\mu$ this analysis offers insight into parametric resurgence.

The spheroidal equation \eqref{eq::SchrodingerSpheroidalEq} reduces to the Mathieu equation when the parameter $\mu$ is set to zero. 
In addition, if the $\cos^2 x $ term is removed from the potential the spheroidal equation is equivalent to the Legendre equation.
The singularities of the potential $V(x) = \cos^2 x + \mu \csc^2 x$ at integer multiples of $\pi$ make the limit $\mu \to 0$ subtle.
This is particularly true for the non-perturbative terms of the transseries and is addressed in section \ref{sec::LimitToMathieu}.

\subsection{Several threads in the literature}
\label{sec::LiteratureReview}
The significance of the spheroidal equation to both mathematics and to physics cannot be understated. 
While, the prolate/oblate spheroidal equations appear in classical mathematical physics---the study of a oscillations of a spheroid\footnote{An ellipsoid where only two axes are of equal length.}---the prolate spheroidal equation surprisingly appears in information/communication theory \cite{doi:10.1073/pnas.2207652119,Slepian}.
When the mathematical foundations of signal processing were being developed C. Shannon sought to refine the following well known statement coming from quantum mechanics: It is not possible to simultaneously concentrate a wave-function in position and momentum spaces.
This led to the following question \cite{doi:10.1073/pnas.2207652119}:
\textit{Suppose you have a band-limited signal---which you observe for some finite time interval. 
What is the best use you can make of your observations?}
This question was answered by Slepian, Landau, and Pollak in the 1960s \cite{Slepian,6773659,6773660}.
At the heart of their work is a certain integral equation---whose solution describes the signal which maximizes the concentration of a band limited signal in the time domain---which has the same eigenfunctions as the prolate spheroidal equation \cite{doi:10.1073/pnas.2207652119,Slepian}.

The results of Slepian, Landau, and Pollak have since been generalized.
For instance, Gr\"unbaum \cite{Grunbaum_2021} has provided a family of integral operators with commuting differential equations.
Further, Gr\"unbaum established a connection between the master symmetries of the KdV hierarchy and this extension of Slepian's result.
Additionally, Longo \cite{Longo:2023ssd} provided an interpretation of the prolate spheroidal operator as an entropy operator, adding clarification to the `lucky accident' that the eigenfunctions of the integral equation found by Slepian, Pollak and Landau are prolate spheroidal functions.

A second remarkable appearance of the spheroidal equation was discovered by A. Connes \cite{connesTraceFormulaNoncommutative1999}.
In \cite{connesTraceFormulaNoncommutative1999} A. Connes gave a spectral interpretation of the zeros of the Riemann zeta function.
The appearance of the spheroidal equation as described in \cite{ConnesMarcolli} is due to the delicate point that obtaining the counting of modes in the corresponding quantum system involves both an ultraviolet and infrared cutoff.
However, because it is impossible to impose a cutoff of both a function and its Fourier transform, Connes utilized the result of Slepian---cutoffs can be approximated using the prolate spheroidal wavefunctions.

This has led to a line of research into the spheroidal equation including \cite{FauvetF.2010Spft,JungRamisFauvetThomann,doi:10.1073/pnas.2123174119,ConnesMarcolli,connes2021spectraltripleszetacycles,connes2024zetazerosprolatewave}.
For example, \cite{FauvetF.2010Spft} sought to answer the challenge posed in \cite{Slepian} of Slepian to find an abstract perspective on the spheroidal wave functions \textit{``that will explain their elegant [properties] in a more natural and profound way."}
The work of Richard-Jung et al \cite{FauvetF.2010Spft} provided a fist step in this direction. 
With a perspective rooted in Riemann-Hilbert theory Richard-Jung et al found new results on the Stokes phenomenon of the spheroidal functions and interpreted them in terms of differential Galois theory.
This was followed by \cite{JungRamisFauvetThomann} where the same authors investigated the eigenvalues of the (prolate) spheroidal equation and provided a novel method for computing them numerically from the zeros of certain polynomials.

We remark that the results presented here are relevant to this line of research.
As will be discussed in section \ref{sec::AllOrdersWKB}, the exact WKB method is closely related to Riemann-Hilbert theory.
Some further relevance is furnished by \cite{doi:10.1073/pnas.2123174119}, in which Connes and Moscovici identified new relationships between the spectrum of the prolate spheroidal operator and the zeros of Riemann zeta function in a semiclassical limit.

Another surprising appearance of the spheroidal equation---this time in physics---comes from bilayer graphene \cite{ChoiMin-Young2011Adot,Hyun_2012}.
In \cite{deGailGraphene} a low energy effective hamiltonian describing the Landau level spectrum of quasiparticles in twisted bilayer graphene was proposed and motivated.
The eigenvalues of the hamiltonian were shown to be expressed exactly in terms of spheroidal eigenvalues \cite{ChoiMin-Young2011Adot}.
Further theoretical grounding for the simplified model used in \cite{ChoiMin-Young2011Adot} has been offered more recently \cite{beckerMagneticResponseProperties2024}.

As noted in \cite{ChoiMin-Young2011Adot} the difference between adjacent Landau levels is controlled by non-perturbative structure in the expansion parameter.
This non-perturbative behavior describes the asymptotic degeneracy found by the decoupling of the two layers of graphene as the twist angle increases.
Within the model used by \cite{ChoiMin-Young2011Adot,deGailGraphene} the non-perturbative structure of the spheroidal eigenvalues encodes the transition between the massive and massless regions of the spectra.
The relevance of the non-perturbative structure of the spheroidal eigenvalues to graphene provides good motivation for a systematic investigation of the non-perturbative properties of the eigenvalues as well as an investigation into various physical contexts in which they may have influence.

Some non-perturbative features of the spheroidal equation have been investigated previously by M\"uller-Kirsten in \cite{Muller+1963+26+48}.
The perturbative and non-perturbative expansions which we compute were chosen to match those computed by M\"uller-Kirsten.
Besides notational differences the analysis provided here uses a slightly different expansion parameter.
We recover M\"uller-Kirsten's expansions in the limit $\mu \propto \hbar$, and in this sense our results carry additional non-perturbative information.

A substantial amount of interest in the spheroidal equation is due to the fact that the spheroidal equation plays a fundamental role in the physics of rotating black holes \cite{TeukolskyThesis}.
In 1973, Teukolsky \cite{TeukolskyThesis} separated the equation describing perturbations of the Kerr metric.
The resulting separated equation is known as the Teukolsky equation.
It describes both perturbations of the Kerr geometry as well as scalar, vector and tensor fields in a Kerr background. 
The Teukolsky equation consists of two parts, a radial equation, and an angular equation.
Together, with proper boundary conditions, they encode the quasinormal mode frequencies of black hole radiation.
The angular equation takes the form of the \textit{spin weighted spheroidal equation}. 
By setting the `spin' to zero, the equation reduces to the spheroidal equation.

Quasinormal modes of black holes are particularly well studied due to the explosion of interest created by the LIGO gravitational wave observations.
That being said, analytic methods have been limited to a small number of techniques.
In particular, two of the most commonly applied are Leaver's continued fraction method \cite{doi:10.1098/rspa.1985.0119,Berti_2009} and the WKB method \cite{PhysRevD.35.3621,Seidel:1989bp,Yang_2013_Branching,Konoplya:2003ii,Konoplya:2019hlu,Konoplya:2023moy,Matyjasek_2017}.

Leaver's continued fraction method \cite{doi:10.1098/rspa.1985.0119} was first applied to the perturbations of the Schwarzschild metric but was adapted to Kerr black holes as well as other geometries.
Notably, it was applied by Berti, Cardoso and Starinets in \cite{Berti_2009} for a number of quasinormal mode calculations.
It functions using a series ansatz to both radial and angular equations which take into account the appropriate boundary conditions.
The WKB method has been applied extensively \cite{PhysRevD.86.104006,Seidel:1989bp,PhysRevD.35.3621,PhysRevD.35.3632,PhysRevD.37.3378}, often including higher order terms in the expansion.
Some examples are \cite{Seidel:1989bp,Konoplya_2003,Konoplya:2019hlu,Matyjasek_2017} which go beyond leading order.
An interesting comparison is to the work of Imaizumi in \cite{Imaizumi::2022,Imaizumi::2023}, who applied the exact WKB method to the computation of quasinormal modes of D3 and M5 branes.
A notable difference comes from Imaizumi's use of summation techniques such as Borel-Pad\'e summation which have been shown (see for instance \cite{Costin:2020hwg}) to improve upon the standard Pad\'e approximant techniques such as those used in \cite{Konoplya_2003,Konoplya:2019hlu,Seidel:1989bp,Matyjasek_2017}.

Imaizumi's analysis fits into a larger context---a recent influx of results on quasinormal modes supplied by the particle physics community \cite{fioravanti2022integrabilitysusysu2matter,fioravanti2022newmethodexactresults,Aminov2020,Bonelli:2021uvf,bianchiMoreSWQNMCorrespondence2022,Aminov:2023jve,Arnaudo:2024rhv,Arnaudo:2024bbd,Hatsuda:2021gtn,Hatsuda_2020,Hatsuda_2021,Silva_2025}.
The connection between these topics is in large part due to the fact that Teukolsky's equation (and therefore the spheroidal equation) is a special case of a confluent Heun equation \cite{Borissov:2009bj,Aminov2020,Bautista:2023sdf}.
A more detailed discussion of some of this research is provided in section \ref{sec::AllOrdersWKB}.

\subsection{Black hole dictionary}
\label{sec::bhDictionary}
For reference, we provide a dictionary connecting the notation used here to the black hole parameters used in the study of quasinormal modes.
The spheroidal equation in the black hole parameters is listed in \cite{Aminov2020}, when in Liouville normal form it is
\begin{equation}\label{eq::BhSpheroidalEq}
  \left(  \derivative[2]{}{\theta}  + \lambda_{lm} + \frac{1}{4} +  \omega^2 a^2 \cos^2 \theta + \left(\frac{1}{4}-m^2 \right) \csc^2\theta\right)S_{m}^l(\theta) =0
\end{equation}
where $a$ is the angular momentum per unit mass of the black hole and ranges between $0$ and $M$. The parameter $M$ is the mass of the black hole.
The parameters $\lambda_{lm}$ are the separation constants.
The labels $m,l$ are the azimuthal and angular harmonic indices.
The quasi-normal mode frequency is given by $\omega$.
The boundary conditions come from a regularity condition at the poles of $\csc \theta$ which can be found in \cite{Aminov2020} or \cite{TeukolskyThesis}.

This equation is equivalent to equation \eqref{eq::SchrodingerSpheroidalEq}.
Specifically, the black hole spheroidal equation \eqref{eq::BhSpheroidalEq} is transformed into equation \eqref{eq::SchrodingerSpheroidalEq} under the identifications
\begin{equation}\label{eq::bhDictionary}
    \begin{aligned}
        &\hbar = \frac{i \sqrt{2}}{a \omega}, & && E = - \frac{1}{a^2 \omega^2} \left( \lambda^{l}_m + \frac{1}{4}\right), && &&&\mu = \frac{1}{a^2 \omega^2} \left(\frac{1}{4}-m^2 \right).
    \end{aligned}
\end{equation}
Let $0<\zeta \leq 1$.
Because the angular momentum per black hole mass must satisfy $0 \leq a \leq M$ the limit $\hbar \to 0$ corresponds to the double limit $a\to \zeta M$ and $M\to \infty$.
This describes a large black hole which is spinning with angular momentum per unit mass $a$ of order of the mass.
This is not the same as the extremal limit.

\section{The exact WKB method}
\label{sec::AllOrdersWKB}

The WKB method can be extended beyond leading order to provide formal series solutions to eigenvalue problems \cite{Dunham_1932}.
These formal asymptotic series are divergent but can be resummed and their monodromy and Stokes phenomenon understood using the framework of exact WKB \cite{Voros::1983,Delabaere1997ExactSE,Bucciotti:2023trp,vanSpaendonck:2023znn}.
In particular, eigenvalues of differential operators can be computed analytically from \textit{exact quantization} conditions.
The building blocks of these conditions are `Voros symbols' or \textit{quantum actions}. 
In section \ref{sec::OverviewOfWKB} we review the standard definition of the quantum actions as well as the Bohr-Sommerfeld condition which allows for calculation of perturbative expansions of eigenvalues.
In section \ref{sec::PNPRelation} we conjecture a form for the first non-perturbative, exponential correction to the perturbative expansion for the energy level.
In section \ref{sec::NumericalTests} we provide a numerical resurgent analysis verifying the proposed non-perturbative asymptotics and computing the Stokes constant.

For the spheroidal equation the exact WKB method is closely related to $SU(n)$, $\mathcal{N}=2$ supersymmetric gauge theory \cite{Grassi_2022,Aminov2020,Yan:2020kkb}. 
More specifically, some Schr\"odinger equations (including the Heun equation and therefore the spheroidal equation) are described by $\mathcal{N}=2$ supersymmetric gauge theories placed in the Nekrasov-Shatashvili limit of the $\Omega$-background.
In this connection, the \textit{classical equation of motion} corresponding to the Schr\"odinger operator appears as the \textit{Seiberg-Witten curve} of the field theory with $\hbar$ playing the role of the $\Omega$-background parameter `$\epsilon_1$.'
The classical period integrals, describing the action of a periodic trajectory, control the masses and central charges of BPS states, while the full series of quantum actions can be computed in terms of the Nekrasov-Shatashvili instanton counting function \cite{Grassi_2022}.
In fact, in \cite{Aminov2020} Aminov, Hatsuda and Grassi used this connection to relate the quasinormal modes of rotating black holes, (and therefore the eigenvalues of the spheroidal equation), to the Nekrasov-Shatashvili instanton counting function.
More details on the connection between exact WKB and $\mathcal{N}=2$ supersymmetric field theories is available in \cite{comanQuantumCurvesTopological2025}.

WKB periods can also be calculated with thermodynamic-Bethe-ansatz (TBA) equations \cite{Imaizumi::2020,Imaizumi::2021,Emery_2021,Ito_2019,Ito2023}.
The TBA approach is perhaps the most efficient approach for calculating quantum periods and solving exact quantization conditions numerically.
The TBA approach however does not produce asymptotic series with full parameter dependencies included.
As described in \cite{Ito_2019}, the seminal work of Voros \cite{Voros::1983} provided a method for computing spectral data from exact quantization conditions using what was termed the ``analytic bootstrap.''
In this approach the WKB periods play a fundamental role and are Borel resummed to form exact quantization conditions.
The discontinuity structure encoded by the Stokes automorphism of the WKB periods determines a Riemann-Hilbert problem which has a solution in terms of a TBA-like system \cite{Ito_2019}.
The connection between TBA and differential equations, the ODE/IM correspondence \cite{Dorey_1999}, has been used for the calculation of quasinormal modes \cite{fioravanti2022newmethodexactresults} of D3 branes and the Reissner-Nordstr\"om geometry.
However, to the knowledge of the author it has not been applied to the case of Kerr black holes.

As a final comment we mention that exact quantization conditions can also be solved analytically in terms of transseries \cite{vanSpaendonck:2023znn}.
Here we lay the foundation for such computations.
For this reason we focus on the computation of the quantum actions exactly and with full parameter dependence included.

\subsection{Quantum actions and the Bohr-Sommerfeld condition}
\label{sec::OverviewOfWKB}
Applying a \textit{WKB} ansatz $\psi(x) = e^{\frac{i}{\hbar}\int^x \phi \dd{x}}$ to the Schr\"odinger equation \eqref{eq::SchrodingerSpheroidalEq} gives the following Riccati equation
\begin{equation}\label{eq::RiccattiEq}
    \phi^2 = 2(E -  \cos^2 x - \mu \csc^2 x) + i \hbar \derivative{}{x}\phi.
\end{equation} 
Setting $\hbar=0$, this equation is algebraic and coincides with the classical equation of motion with the \textit{action density} $\phi$ playing the role of momentum.
In the complex domain the classical equation of motion defines a Riemann surface which happens to be a torus\footnote{The classical equation defines a two parameter family of tori.}, call it $\Sigma$.

The equation \eqref{eq::RiccattiEq} has an asymptotic series solution \cite{Dunham_1932} for the action density $\phi$.
Let
\begin{equation}
    \phi(\mu,E,\hbar ,x) = \sum_{k=0}^\infty \phi_k(\mu ,E,x) \hbar^k.
\end{equation}
Each of the $\phi_k \dd{x} $ are differentials on the Riemann surface $\Sigma$.
More precisely they are differentials on $\Sigma$ where the points $x= \dots ,-2\pi ,-\pi,0,\pi , \dots$ (which correspond to simple poles of the $\phi_k$) are removed.\footnote{It is important to point out that the series has the property that the odd terms are all total derivatives with the exception of $\phi_1$.}
This means that we can integrate them along paths on $\Sigma$ and define quantum actions.

\paragraph*{The Quantum Actions.}
Let $\gamma$ be a cycle on $\Sigma$ then the \textit{quantum action} along $\gamma$ is defined to be
\begin{equation}
    N_\gamma (E,\mu,\hbar) = \sum_{k=0}^\infty \left( \oint_\gamma \phi_k(\mu,E,x) \dd x \right) \hbar^k.
\end{equation}
Because $\Sigma$ is a torus there are two independent cycles.
Call them $A,B$ and choose $A$ so that it corresponds to a classical periodic orbit.
One way to identify $A$ and $B$ is to construct $\Sigma $ using the `cut-and-paste' method.
In this perspective the $A$ cycle should be chosen to move around the classical turning points associated to a well of the potential.
These turning points coincide with ramification points of $\Sigma$ which the `cuts' must connect.     

\paragraph{The Bohr-Sommerfeld Condition.}\label{conj::BohrSommerfeld}
Let $l \in \N_0 $.
The solution $E_A(l,\mu,\hbar)$ to the following equation
\begin{equation}\label{eq::BohrSommerfeld}
    2\pi \hbar l = N_A(E_A, \mu, \hbar)
\end{equation}
is an asymptotic expansion for the $l$th eigenvalue of the Schr\"odinger equation in the limit $\hbar \to 0$ and  $l \to \infty $ such that the limit of the product $\hbar l$ is fixed and proportional to the classical action.
The limit is analogous to a t'Hooft double scaling limit \cite{LoChiatto:2023bam}.

The low energy expansion of the quantum actions can be used to generate all orders of the Rayleigh-Schr\"odinger perturbative series associated to $A$.
In particular, the perturbative series for the energy is determined by calculating the `low energy' expansion of the quantum action $ N_A(E, \mu, \hbar)$ and then by solving the Bohr-Sommerfeld condition for $E_A$ using series inversion. 
After calculating the quantum action in section \ref{sec::TheQuantumActions} we use this approach to find the perturbative series for $E_A(l,\mu,\hbar)$ in section \ref{sec::PertEnergyLevel}.

\subsection{The classical actions}
\label{sec::ClassicalActions}

At the classical level the actions are given by integrals of the form
\begin{equation}\label{eq::Classical action}
    N_{0,\gamma} (E,\mu) =  \int_\gamma \sqrt{2(E - \cos^2 x -\mu \csc^2 x )} \dd x 
\end{equation}
where $\gamma$ is a cycle on the Riemann surface $\Sigma$.
For each $\gamma$ the classical actions can be expressed in terms of elliptic integrals of the first, second and third kinds.
The elliptic integrals are denoted by $ \mathbf{K}, \mathbf{E}, \mathbf{\Pi}$ the complete elliptic integrals of the first, second and third kinds, respectively.
To fix notation they are defined in appendix \ref{sec::EllipticIntegrals}.
The integral \eqref{eq::Classical action} can be expressed in terms of elliptic integrals with the following steps:
    Let $a = \frac{1}{2}(E+1 - \sqrt{(E-1)^2 +4 \mu}) $ and $b= \frac{1}{2}(E+1 + \sqrt{(E-1)^2 +4 \mu})$.
    The change of coordinates $x\mapsto \arccos(\sqrt{a} X)$ transforms the integral into the following
    \begin{equation}\label{eq::EllipticIntegral}
      N_{0,\gamma} (E,\mu) = a \sqrt{2 b} \int_{\gamma'} \frac{1}{1- a z^2} \sqrt{(1-X^2)\left(1-\frac{a}{b} X^2 \right)} \dd X.
    \end{equation}  
    The integral \eqref{eq::EllipticIntegral} can be computed by calculating the antiderivative of the integrand using a computer algebra system such as Mathematica.
    The cycles of integration corresponding to the action and dual action can be identified by considering the images of the turning points.
    In particular, we find that the cycle determining the action circles the branch points $X = -1$ and $X=1$ while the dual action is determined by the trajectory moving from $X=1$ to $X=\sqrt{b/a}$.
    Combining these considerations provides expressions for the action and dual action.
    With the notation $k = \sqrt{a/b}$, the action is given by 
\begin{equation}\label{eq::EllipticIntegralRepresentation}
    \begin{aligned}
        N_{0,A} (E,\mu) = \frac{4\sqrt{2}}{k\sqrt{a}} \left(   k^2 (a-1) \mathbf{K}\left(k \right)+ (a-1)\left(a-k^2\right) \mathbf{\Pi} \left(a , k \right) +a k \,\mathbf{E}\left( k \right)  \right).
    \end{aligned}
\end{equation}
    The dual action can be expressed as follows
\begin{equation}
  \begin{aligned}
      N_{0,B}(E,\mu) &=  \frac{2\sqrt{2} }{\sqrt{a} k^2 }\Bigg(  \left(a-k^2\right) (a-1) \mathbf{\Pi} \left(b, \frac{1}{k}\right)+ \left(a-k^2\right) \mathbf{K}\left(\frac{1}{k}\right)
      \\ & +a k^2 \mathbf{E}\left(\frac{1}{k}\right) \Bigg) -\frac{1}{2} N_{0,A}(E,\mu) .
  \end{aligned}
\end{equation}

\begin{figure}\begin{center}
    \includegraphics[width = 10cm]{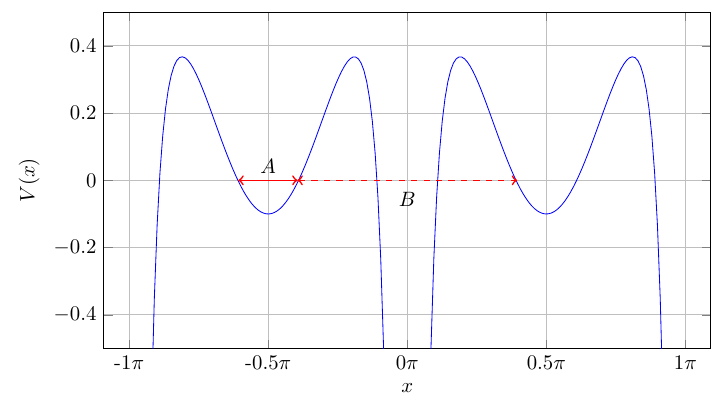}
\end{center}
\caption{The figure shows the potential in the region $-1<\mu <0$ with the $A$ and $B$ cycles sketched by the solid and dashed arrows respectively.
The harmonic well defining the $A$ cycle is separated from an infinitely deep well associated to a pole by a finite barrier.
\label{fig::PotentialFigure}}
\end{figure}

The definition of the $A$ and $B$ cycles depends on the value of the parameters $\mu$ and $E$, as these determine both the shape of the potential as well as the classical and non-classical regions.
Here we focus on the region of parameters $-1<\mu<0$ and $\mu<E< 1 - 2 \sqrt{|\mu|}$.
In this case the harmonic well defining the $A$ cycle is separated from an infinitely deep well associated to a pole of the potential by two finite barriers on either side, see figure \ref{fig::PotentialFigure}.
On one hand, it might be expected that the $B$ cycle describes the tunneling rate out of the harmonic well and into the infinitely deep well.
On the other, the $B$ cycle could be determined by a trajectory which tunnels between neighboring harmonic wells.
The difference between the choices is an overall factor of two as well as a contribution coming from the residue of the pole.

Numerical tests show that the second interpretation `tunneling between harmonic wells' is responsible for the leading non-perturbative behavior.
This has the interesting consequence that the leading non-perturbative structure of the spheroidal eigenvalue in this region reduces to that of the Mathieu equation in the limit $\mu \to 0$.

\subsection{Picard-Fuchs differential operators}
\label{sec::PFEs}
The WKB expansion is intimately tied to the geometry of the Riemann surface $\Sigma$.
In particular, variations of the parameters $(\mu,E)$ lead to deformations of $\Sigma$.
However, for each pair $(\mu,E) \in \C$ (away from critical points) the Riemann surfaces are diffeomorphic.
The deformation is therefore not `topological,' but occurs at the level of the underlying complex/Hodge structure.

Varying the parameters leads to the `monodromy of homology' described by the Picard-Lefschetz formula \cite{Hwa_Hua_Teplitz_1966}.
This monodromy leads to a monodromy of the integrals defining the quantum actions and is closely related to the Stokes phenomena described by the Delabaere-Dilinger-Pham formula \cite{Delabaere1997ExactSE}.
The deformation of the integrals is captured by a locally flat connection, the Gauss-Manin connection, which describes the parallel transport of de-Rham classes (and therefore period integrals) as the parameters are varied.
The Gauss-Manin connection determines a system of differential equations, known as Picard-Fuchs equations, satisfied by the period integrals on $\Sigma$.

A differential $\phi$ on $\Sigma$ defines a de-Rham cohomology class $[\phi]$.
Differentiating $\phi$ with respect to the external parameters $E, \mu$ using the Gauss-Manin connection produces new classes $\pdv{\phi}{E},\pdv{\phi}{\mu},\pdv[2]{\phi}{\mu}{E},\dots$.
The geometric fact that the de-Rham cohomology group is finitely generated, with rank equal to twice the genus of $\Sigma$, implies that after a certain number of derivatives  with respect to the parameters $E$ and $\mu$ the classes must be linearly dependent.
Upon integration (along a cycle on $\Sigma$) these linear dependencies lead to Picard-Fuchs equations.

A utility and insight of this picture to WKB is to the computation of quantum actions.
In particular, the quantum actions can be determined from Picard-Fuchs differential operators. 
They take the form of differential equations satisfied by the classical actions as well as differential operators which express quantum actions entirely in terms of the classical action.
Approaches to computing Picard-Fuchs equations for WKB periods can be found in \cite{Basar_2017,Raman_Subramanian_2020,Fischbach_Klemm_Nega_2019,Ito_2019,Ito2023}, here the Picard-Fuchs differential operators were calculated using a modification of the method described in \cite{Meynig:2024nam} to the trigonometric potential of the spheroidal equation.

\paragraph*{Classical Picard-Fuchs equations.}
Let $T = \left(E-\mu\right) \left(4 \mu+\left(E-1\right)^{2}\right). $
The classical action $N_{0}(E,\mu)$ of equation \eqref{eq::Classical action} associated to the spheroidal equation satisfies the following Picard-Fuchs equations:
\begin{equation}\label{eq::ClassicalPFeq}
    \begin{aligned}
    \frac{\partial^2 N_0 }{ \partial E^2 } &=  \frac{1}{4T}\left(  -E+2 \mu+1  \right) N_0   - \frac{\mu}{2T} \left(1+E\right) \frac{\partial N_0 }{ \partial E }  
      + \frac{\mu}{2T} \left(E-2 \mu-1\right)  \frac{\partial N_0 }{ \partial \mu } ,
    \\    \frac{\partial^2 N_0 }{ \partial E \partial \mu } &=   -\frac{1}{4T}\left( 1 + E   \right) N_0  + \frac{1}{2T}\left(  E^{2}- E+2 \mu \right) \frac{\partial N_0 }{ \partial E } 
      + \frac{\mu}{2T} \left(E+1\right) \frac{\partial N_0 }{ \partial \mu },
    \\  \frac{\partial^2 N_0 }{  \partial \mu^2 } &=  \left( \frac{E^{2}-E+2 \mu}{4 \mu T}  \right) N_0 + \left( -\frac{1}{2\mu}+\frac{-E^{2}+E-2 \mu}{2T } \right) \frac{\partial N_0 }{ \partial E } 
    \\ & + \left( \frac{2 E^{2}-2 E+4 \mu}{4 T} -\frac{1}{2\mu} \right) \frac{\partial N_0 }{ \partial \mu }.
    \end{aligned}
\end{equation} 
The above system of Picard-Fuchs equations follows from the differential equations satisfied by $\mathbf{K}, \mathbf{E}, \mathbf{\Pi}$ as well as equation \eqref{eq::EllipticIntegralRepresentation}.
We emphasize that the differential equations are independent of cycle. 
The differential equations satisfied by the elliptic integrals are listed in appendix \ref{sec::EllipticIntegrals}. 

Series expansions for the classical action and dual action are accessible by solving the Picard-Fuchs equations with series ansatz. 
This is particularly useful in the case of the dual action where the necessary series for the elliptic integrals are more difficult to access.
The difficulty is due to the appearance of logarithms in the expansions.
The series are given by the following:
Let $C$ be the cycle which circles the pole at $x=0$ of $\phi_0$. 
Near $E= \mu $ the classical actions are given by 
\begin{equation}\label{eq::ClassicalActions}
    \begin{aligned}
        N_{0,A}  &=  \pi \frac{\sqrt{2} (E-\mu)}{\sqrt{\mu+1}} +   \pi  \frac{(1-2 \mu) (E-\mu)^2}{4 \sqrt{2} (\mu+1)^{5/2}} +  \pi  \frac{(8 (\mu-3) \mu+3) (E-\mu)^3}{32 \sqrt{2} (\mu+1)^{9/2}}
        \\ & +  \pi \frac{\left(25-10 \mu \left(8 \mu^2-60 \mu+45\right)\right) (E-\mu)^4}{512 \sqrt{2} (\mu+1)^{13/2}}  + \dots ,
    \\  N_{0,B} &=  i   2\sqrt{2} \left(\sqrt{\mu+1}-\sqrt{\mu} \text{ arcsinh}\left(\sqrt{\mu}\right)\right) +  \pi\sqrt{2 \mu } -  \frac{i (18 \mu-5) (E-\mu)^2}{16 \sqrt{2} (\mu+1)^{5/2}} 
    \\ & + \frac{i \left(48 \mu^2-96 \mu+9\right) (E-\mu)^3}{64 \sqrt{2} (\mu+1)^{9/2}} + \dots + \frac{i}{2\pi}\left( 1 - \log\left(\frac{E-\mu}{16 (\mu+1)^2}\right) \right) N_{A,0} ,
    \\  N_{0,C} &= \pi\sqrt{2\mu} . 
    \end{aligned}
\end{equation}

\subsection{The quantum actions}
\label{sec::TheQuantumActions}
As mentioned above, the method described in \cite{Meynig:2024nam} was used to generate the Picard-Fuchs equations.
To compute the quantum actions we first find normal forms for the cohomology class of $\pdv{\phi}{E}$ using the Griffiths-Dwork reduction algorithm.
The Griffiths-Dwork algorithm is described in \cite{GriffithsI,Griffiths:doi:10.1073/pnas.55.5.1303,Cox_Katz_2000,Lairez_2015,Muller-Stach_Weinzierl_Zayadeh_2014,Morrison:1991cd} and has been used in the analysis of Feynman integrals \cite{Muller-Stach_Weinzierl_Zayadeh_2014,lairezAlgorithmsMinimalPicard2023} as well as in integrals appearing in mirror symmetry.

The Griffiths-Dwork reduction applies to rational integrals (multiple integrals of rational functions).
However, the integrals appearing in quantum actions are not rational integrals but are the \textit{residues} of rational integrals. 
The rational integrals are related to the quantum action by the formula 
\begin{equation}
  \int_{\gamma} \dv{\phi}{E}\dd x = \int_{\gamma} \text{res}\, R \dd{x} \dd{p} 
\end{equation}
where $R$ is the symbol of the resolvent $(\hat H  -E )^{-1}$ where $\hat{H}$ is the Schr\"odinger operator defined by \eqref{eq::SchrodingerSpheroidalEq}.
The multivariable residue map is described in \cite{Meynig:2024nam,GriffithsI,Griffiths:doi:10.1073/pnas.55.5.1303}.
The symbol of the resolvent $R$ can be generated from the small $\hbar$ series expansion of
\begin{equation}
   R =  \frac{1}{1- \frac{\hbar}{H-E} \left( i p \pdv{}{x} - \frac{\hbar}{2} \pdv[2]{}{x}  \right)} \frac{1}{H-E}
\end{equation}
with $H = \frac{1}{2} p^2 + \cos^2 x + \mu \csc^2 x $ the classical hamiltonian \cite{Meynig:2024nam}.
The mapping 
\begin{equation}\label{eq::CoordinateTransformation}
    (x,p) \mapsto \left( \arccos(X), -\frac{P}{\sqrt{1-X^2}} \right)
\end{equation} 
converts the terms in the power series expansion of $R$ to rational functions of $(X,P)$. 
For example, the classical equation of motion $H = E$ corresponds to the quartic elliptic curve
\begin{equation}
    \frac{1}{2} P^{2} -  X^{4} +\left(1 + E \right) X^{2}-E+\mu  = 0.
\end{equation}

Once in the form of a power series of rational functions the Griffiths-Dwork algorithm computes normal forms of the quantum periods efficiently.
The quantum period is given by a series in even powers of $\hbar$ of the form
\begin{equation}\label{eq::Ederivative}
    \oint_{\gamma} \dv{\phi}{E} \dd x = \sum_{n=0}^\infty \Pi_{2n}(E,\mu) \hbar^{2n}
\end{equation}
where $\Pi_0$ is the classical period associated to the cycle $\gamma$ on the Riemann surface $\Sigma$.
The algorithm outputs relations of the form 
\begin{equation}\label{eq::PeriodRelation}
    \Pi_{2n} = \alpha_{2n} (E,\mu) \Pi_0 + \beta_{2n} (E,\mu) \pdv{\Pi_0}{E} .
\end{equation}
These relations can be converted to a relation for the quantum actions $N_{2n}(E,\mu)$ through integration with respect to $E$.
This integration is non-trivial due to the appearance of simple poles in the action density $\phi$.
Taking the $E$-derivative as in \eqref{eq::Ederivative} removes these poles, allowing for a straight-forward application of the Griffiths-Dwork algorithm.

Integration of the period relations \eqref{eq::PeriodRelation} is possible by first guessing that the integrated relations are of the form
\begin{equation}\label{eq::FormOfActionRelation}
   N_{2n} = a_{2 n}(E,\mu)\Pi_0 + b_{2n}(E,\mu) \pdv{\Pi_0}{E} + c_{2n}(\mu) \pdv{N_0}{\mu}
\end{equation}
Differentiating \eqref{eq::FormOfActionRelation} using the Picard-Fuchs equations \eqref{eq::ClassicalPFeq} to eliminate $\pdv{N_0}{\mu}{E}$ and $\pdv[3]{N_0}{E}$ gives coupled ordinary differential equations for $\derivative{a_{2 n}}{E},\derivative{b_{2 n}}{E}$ in terms of $\alpha_{2n} ,\beta_{2n}$ and $c_{2n}$.
We find the particular solution to each of these equations using a symbolic software such as in Maple.
Solving for the particular solution requires additonal input, the function $c_{2n}(\mu)$.
It is fixed by the simple pole of the action density $\phi$. 
Because both $\pdv{N_0}{\mu}{E}$ and $\pdv{N_0}{E}$ have no simple poles, each $c_{2n}(\mu)$ must be entirely determined by the residue of the pole.
Near $x=0$ the potential $V(x)= \cos^2 x + \mu \csc^2 x$ is of the form
\begin{equation}
    V(x) = \frac{\mu }{x^2} +\left(\frac{\mu }{3}+1\right)+\left(\frac{\mu }{15}-1\right) x^2 + \dots
\end{equation}
We have that near a pole of the potential the action density $\phi$ has a Laurent expansion of the form,
\begin{equation}
    \phi(\mu,E,\hbar,x) = \frac{b_{-1}}{x}+ \text{o}(1) .
\end{equation}
The coefficient $b_{-1}$ can be calculated directly from equation \eqref{eq::RiccattiEq}.
That is, the equation implies
\begin{equation}
    \begin{aligned}
        \frac{b_{-1}}{x} +\text{o}(1)  &= \sqrt{2(E - V(x))+i \hbar \frac{d}{dx}\phi} \\
            & = \frac{1}{x} \sqrt{-2\mu + i \hbar x^2 \frac{d}{dx} \frac{b_{-1}}{x}} + \text{o}(1)  \\
            & = \frac{i}{x} \sqrt{ 2\mu + i \hbar b_{-1} } +\text{o}(1) .
    \end{aligned}
\end{equation}
Solving for $b_{-1}$ gives
\begin{equation}
    b_{-1} =   \pm i\sqrt{ 2\mu + \frac{1}{4}\hbar^2} - \frac{i \hbar }{2}
\end{equation}
and suggests the quantum action has a simple pole at $x=0$ with residue $b_{-1}$.
For additional discussion of this result in the context of more general potentials see \cite{Cavusoglu:2023bai}.
From the residue calculation we determine
\begin{equation}
    c_{2k}(\mu) = 4 \binom{\frac{1}{2}}{k} \left(\frac{1}{8 \mu }\right)^k.
\end{equation}

Finally, using the first Picard-Fuchs equation in \eqref{eq::ClassicalPFeq} the relations of the form \eqref{eq::FormOfActionRelation} can be converted to a relation of the form
\begin{equation}
    N_{2n} = r_{2n}(E,\mu) N_{0} + s_{2n}(E,\mu)  \pdv{N_0}{E} +  t_{2n}(E,\mu)  \pdv{N_0}{\mu} . 
\end{equation}
These relations determine a differential operator $\mathcal{L}_{E,\mu}$ which can be used to calculate the quantum action and dual action as series in $\hbar$ purely from the classical quantities.
Up to $\hbar^4$ the differential operator output by the described calculation is given by
\begin{equation}\label{eq::DiffOp}
    \begin{aligned}
       \mathcal{L}_{E,\mu}  &=  1 +  \frac{\hbar^{2}}{12} \Bigg( \frac{1}{4}\frac{1}{ \mu - E }+  \frac{1-E}{\left(E-1\right)^{2}+4\mu} 
        +\left(1 - \frac{1}{2}\frac{E}{\mu - E} + \frac{(E-1)^{2}}{\left(E-1\right)^{2}+4\mu}\right) \frac{\partial}{\partial E} 
       \\ & +  \left(\frac{3}{2} + 2 \frac{\mu \left(E-1\right)}{ \left(E-1\right)^{2}+4\mu}+\frac{\mu}{2 \left(E-\mu\right)} \right) \frac{\partial}{\partial \mu} \Bigg)+\mathrm{O}\! \left(\hbar^{4}\right).
    \end{aligned}
\end{equation}
Let $N_0$ be the classical action associated to any cycle on $\Sigma$.
Then the differential operator $\mathcal{L}_{E,\mu}$ satisfies
\begin{equation}\label{eq::QuantumActions}
    \begin{aligned}
        &N(E,\mu,\hbar)  =   \mathcal {L}_{E,\mu} N_0(E,\mu)  +\pi i \hbar.
    \end{aligned}
\end{equation}
The addition of the term $\pi i \hbar$ comes from the integral of $\phi_1$.
The differential operator $\mathcal {L}_{E,\mu} $ has been calculated to $\mathcal{O}(\hbar^8)$ and is available in the Appendix.

\subsection{The perturbative energy level}
\label{sec::PertEnergyLevel}

The perturbative energy level can be determined using the Bohr-Sommerfeld conditon and the quantum action.
Solving the Bohr-Sommerfeld condition \ref{eq::BohrSommerfeld} with Lagrange inversion gives the following
\begin{equation}\label{eq::PerturbativeSeries}
    \begin{aligned}
        &E_A(l,\mu,\hbar) = \sum_{k=0}c^{(0)}_k(l,\mu) \hbar^k 
        \\ &=\mu + \left(l+\frac{1}{2}\right) \sqrt{2}\, \sqrt{\mu+1}\, \hbar +\frac{\left(2 \mu-1\right) \left(4 \left(l+\frac{1}{2}\right)^{2}+1\right) \hbar^{2}}{16 \mu+16}
        \\ & +\frac{\left(l+\frac{1}{2}\right) \sqrt{2}\, \left(64 \mu \left(l+\frac{1}{2}\right)^{2}+8 \mu^{2}-4 \left(l+\frac{1}{2}\right)^{2}+72 \mu-3\right) \hbar^{3}}{128 \left(\mu+1\right)^{\frac{5}{2}}} 
        \\ & +\Bigg(1242 \mu -9 -288 \mu ^2+\left(-2560 \mu ^2+3360 \mu -80\right) \left(l+\frac{1}{2}\right)^4
        \\ & +\left(-4352 \mu ^2+9168 \mu -136\right) \left(l+\frac{1}{2}\right)^2\Bigg) \frac{\hbar^4}{4096 (\mu+1)^4}
        \\ & +\Bigg( (-128 \mu ^4+14592 \mu ^3-166656 \mu ^2+123360 \mu -405) \left(l +\frac{1}{2}\right)
        \\ & + (71680 \mu ^3-416640 \mu ^2+223360 \mu -1640)\left(l +\frac{1}{2}\right)^3
        \\ & + 48 (512 \mu ^3-2144 \mu ^2+896 \mu -11)\left(l +\frac{1}{2}\right)^5
        \Bigg)\frac{\hbar^5}{32768 \sqrt{2} (\mu +1)^{11/2}} + ...
    \end{aligned}
\end{equation}
The above expression can be checked using the `BenderWu' Mathematica package \cite{Sulejmanpasic_Unsal_2016}.
In addition it can be verified by comparing to the series reported by M\"uller-Kirsten in \cite{Muller+1963+26+48} with the appropriate notational changes.

\section{The P/NP relation}
\label{sec::PNPRelation}

While the perturbative energy level is determined by the Bohr-Sommerfeld condition, the geometry of the Riemann surface $\Sigma$ produces an additional constraint.
This constraint comes from Riemann's bilinear identity. 
In the context of quantum mechanics, it relates the perturbative and non-perturbative properties of the spectrum.
As demonstrated in \cite{vanSpaendonck:2023znn,Basar:2015xna}, quantum bilinear identities encode P/NP relations.
Additional context for P/NP relations comes from \cite{GabrielAlvarez2000,vanSpaendonck:2023znn,Basar_2017,Cavusoglu:2023bai,Kemalresurgencedeformedgenus1curves}.

\paragraph*{Riemann's Bilinear identity.}
Consider a Riemann surface $X$ of genus $g$.
Let $\omega_1,\omega_2$ be two (meromorphic) differentials on the Riemann surface $X$.
There is a symplectic basis of cycles $A_i,B_i$ on the Riemann surface $X$ such that
\begin{equation}\label{eq::RiemannsBilinearIdentity}
    \int_\Sigma \omega_1 \wedge  \omega_2 = \sum_{i=1}^g \left(\int_{A_i} \omega_1 \int_{B_i} \omega_2 - \int_{B_i} \omega_1 \int_{A_i} \omega_2 \right)
\end{equation}
which is Riemann's bilinear identity \cite{eynard2018lecturesnotescompactriemann,Gorsky_2015}.

The left-hand-side of the equation becomes a residue integral \cite{eynard2018lecturesnotescompactriemann,vanSpaendonck:2023znn}.
For instance, the bilinear identity studied in \cite{Basar_2017} which describes the action and dual action of the Mathieu potential has a left-hand-side given (in the appropriate normalization) by $4\pi$.
This is especially relevant to the spheroidal equation as the actions satisfy a bilinear relation which reduces to that of the Mathieu equation in the limit $\mu \to 0$.

The importance of bilinear relations to spectra has been understood through the lens of supersymmetric gauge theories \cite{Gorsky_2015,vanSpaendonck:2023znn,Basar:2015xna}.
In this perspective the bilinear identity comes from the Matone relation \cite{Matone_1995}.
Within the TBA framework the quantum bilinear identity can be understood as ``an effective central charge'' \cite{Imaizumi::2020,Ito_2019}.
A complementary approach is to view the bilinear identity as a Wronskian condition coming from a system of Picard-Fuchs equations. 
This perspective is developed in section \ref{sec::VerifyBilinear}.

\subsection{The quantum bilinear identity}
\label{sec::VerifyBilinear}

Let $N_A, N_B$ be the quantum action and dual action of the spheroidal equation.
The quantum bilinear identity is
\begin{equation}\label{eq::SpheroidalBilinear}
    4 i \pi =  \left( N_B -2\mu \frac{\partial N_B}{\partial \mu} - \hbar \frac{\partial N_B}{\partial \hbar} \right) \frac{\partial N_A}{\partial E} - \left( N_A -2\mu \frac{\partial N_A}{\partial \mu} - \hbar \frac{\partial N_A}{\partial \hbar} \right) \frac{\partial N_B}{\partial E} .
\end{equation}
This section addresses the origins of this \textit{quantum} bilinear identity.

A novel approach to the bilinear relation for the spheroidal equation comes by considering it as a generalized `Wronskian' relation.
This complements the existing approaches found in the literature \cite{Cavusoglu:2023bai,Kemalresurgencedeformedgenus1curves,Dunne_2014,vanSpaendonck:2023znn,marshakov1998seibergwitten}.
The calculation involves `geometric' input, systems of Picard-Fuchs equations here given by equation \eqref{eq::ClassicalPFeq}.
By taking derivatives of the right hand side of the bilinear identity and using the Picard-Fuchs equations to reduce higher order derivatives we find that the quantity on the left-hand-side of the equation must be constant.
The constant is then fixed by the leading terms of the action and dual action.
In order, to apply the Wronskian approach it is convenient to look at the classical structure first.

\paragraph*{The classical bilinear identity.}
It is useful to reorganize the data of the Picard-Fuchs equations \eqref{eq::ClassicalPFeq}.
The action $N_0$ and its derivatives $\partial_E N_0,\partial_\mu N_0$ span a vector space and it makes sense to map linear combinations of them to column vectors.
That is,
\begin{equation}
    v_0 N_0 + v_1 \partial_E N_0 + v_2 \partial_\mu N_0  \mapsto \begin{pmatrix}
        v_0 \\
        v_1 \\ 
        v_2
    \end{pmatrix}.
\end{equation}

In this notation the data of the Picard-Fuchs equations is encoded in a matrix valued one form $U= U_E \dd{E} + U_\mu \dd{\mu} $ constructed such that for each vector $\vec v$ we have that
\begin{equation}\label{eq::Udef}
  \dd (v_0 N_0 + v_1 \partial_E N + v_2 \partial_\mu N_0) \mapsto \dd \vec v +U \vec v.
\end{equation}
The coefficients of $U$ can be `read-off' from the Picard-Fuchs equations.
The columns of $U$ `are' Picard-Fuchs equations. 
The above equation motivates the definition of  
\begin{equation}
      \nabla_{GM} = \dd + U\wedge
\end{equation}
which is an example of a construction known as the Gauss-Manin connection, a `covariant derivative' which encodes the parallel transport of period integrals with respect to the space of parameters.
    
We define the following $3\times 3 $ matrices:
\begin{equation}\label{eq::GMC}\setlength\arraycolsep{6pt}
    \begin{aligned}
  &U_E = \begin{pmatrix}
        0 & \frac{-E+2 \mu+1}{4 \left(E-\mu\right) \left(4 \mu+\left(E-1\right)^{2}\right)} & \frac{-1-E}{4 \left(4 \mu+\left(E-1\right)^{2}\right) \left(E-\mu\right)} 
        \\
         1 & -\frac{\mu \left(1+E\right)}{2 \left(E-\mu\right) \left(4 \mu+\left(E-1\right)^{2}\right)} & \frac{2 E^{2}-2 E+4 \mu}{4 \left(4 \mu+\left(E-1\right)^{2}\right) \left(E-\mu\right)} 
        \\
         0 & \frac{\mu \left(E-2 \mu-1\right)}{2 \left(E-\mu\right) \left(4 \mu+\left(E-1\right)^{2}\right)} & \frac{\mu \left(1+E\right)}{2 \left(4 \mu+\left(E-1\right)^{2}\right)\left(E-\mu\right)} 
        \end{pmatrix},
 \\ &U_\mu = \begin{pmatrix}
        0 & \frac{-1-E}{4 \left(4 \mu+\left(E-1\right)^{2}\right) \left(E-\mu\right)} & \frac{E^{2}-E+2 \mu}{4 \mu \left(4 \mu+\left(E-1\right)^{2}\right) \left(E-\mu\right)} 
        \\
         0 & \frac{ E^{2}- E+2 \mu}{2 \left(4 \mu+\left(E-1\right)^{2}\right) \left(E-\mu\right)} & -\frac{1}{2\mu}+\frac{-E^{2}+3E-4 \mu}{2 \left(4 \mu+\left(E-1\right)^{2}\right) \left(E-\mu\right)}
        \\
         1 & \frac{\mu \left(1+E\right)}{2 \left(4 \mu+\left(E-1\right)^{2}\right) \left(E-\mu\right)} & -\frac{1}{2\mu} - \frac{ E^{2}- E + 2 \mu}{2 \left(2 \mu+\left(E-1\right)^{2}\right) \left(E-\mu\right)}
        \end{pmatrix}.
    \end{aligned}
\end{equation}
The connection $U$ satisfying \eqref{eq::Udef} and corresponding to the Picard-Fuchs equations in \eqref{eq::ClassicalPFeq} is determined by
\begin{equation}
    U = U_E \dd E + U_\mu \dd \mu .
\end{equation}

Let $N_{0,A},N_{0,B}$ and $ N_{0,C}$ be the general solutions to the classical Picard-Fuchs equations given by equation \eqref{eq::ClassicalPFeq}.
In particular, let $N_{0,C} = \sqrt{2\mu}$ be the solution corresponding to the pole.
We define another $3\times 3$ matrices:
\begin{equation}\setlength\arraycolsep{6pt}
    M =   \begin{pmatrix}
        N_{0,A} &  \partial_E N_{0,A}  &  \partial_\mu N_{0,A} \\
        N_{0,B} &  \partial_E N_{0,B}  &  \partial_\mu N_{0,B} \\
        N_{0,C} &  \partial_E N_{0,C}  &  \partial_\mu N_{0,C} \\
      \end{pmatrix}.
\end{equation}
Then evaluating the determinant with $N_{0,C} =\sqrt{2\mu}$ shows that
\begin{equation}
    \sqrt{2\mu} \det M = \left( N_{0,A} -2\mu \frac{\partial N_{0,A}}{\partial \mu} \right) \frac{\partial N_{0,B}}{\partial E} - \left( N_{0,B} -2\mu \frac{\partial N_{0,B}}{\partial \mu}\right) \frac{\partial N_{0,A}}{\partial E}.
\end{equation}
We have that
\begin{equation}
    \dd M = MU
\end{equation}
which is evident by evaluating the left hand side using the Picard-Fuchs equations and comparing to the right hand side.
Evaluating the differential of the determinant using `Jacobi's formula' gives
\begin{equation}
    \dd \det M  = \det M \tr (M^{-1} \dd M ) = \det M \tr (M^{-1} M U ) =\det M \tr U.
\end{equation}
From equation \eqref{eq::GMC} the trace of $U$ is given by $\tr U = - \frac{1}{2\mu} \dd \mu$.
Therefore solving the differential equation gives 
\begin{equation}
        \det M = e^{ \int \tr U}=\frac{c}{\sqrt{\mu}}
\end{equation}
where $c$ is a constant of integration.
Multiplying both sides of the above equation by $\sqrt{\mu}$ gives
\begin{equation}
    \sqrt{\mu} \det M  = c .
\end{equation}

\paragraph*{The quantum bilinear relation.}
The quantum bilinear relation \eqref{eq::SpheroidalBilinear} can be verified order by order in $\hbar$ using the Wronskian approach.
The approach is simplified by the fact that the differential operator $\mathcal{L}_{E,\mu}$ defined in equation \eqref{eq::DiffOp} can be used to construct a change of basis from the classical quantities $\{ N_0 ,\partial_E N_0,\partial_\mu N_0 \}$ to the quantum analogs $\{ N ,\partial_E N,\partial_\mu N \}$.
However, the quantum action includes a contribution from the integral of $\phi_1$ which defines the \textit{Maslov correction} and cannot be described in the classical basis of periods.
To account for it, it is convenient to construct a `quantum basis' spanned by $\{N,\partial_E N ,\partial_\mu N ,\partial_\hbar N \}$.

It is useful to consider the $\hbar \to 0$ limit of the `quantum basis.'
Let $\hbar \nu$ be the Maslov correction. 
Then 
\begin{equation}
    \lim_{\hbar \to 0} \{N,\partial_E N ,\partial_\mu N ,\partial_\hbar N \} = \{N_0, \partial_E N_0, \partial_\mu N_0 , \nu \}.
\end{equation}
This motivates the mapping defined by
\begin{equation}\label{eq::defFourVecs}
    v_0 N_0 + v_1 \partial_E N_0 + v_2 \partial_\mu N_0 +  v_3 \nu \mapsto 
    \begin{pmatrix}
        v_0 \\
        v_1 \\
        v_2 \\
        v_3
    \end{pmatrix}.
\end{equation}

    Let $U_E$ and $U_\mu$ be the $3\times 3$ matrices introduced in equation \ref{eq::GMC}.
    Let $\varphi $ be the map defined by equation \eqref{eq::defFourVecs}.
    Let $\vec v$ be given by
    \begin{equation}
        \vec v = \varphi \left( v_0 N_0 + v_1 \partial_E N_0 + v_2 \partial_\mu N_0 +  v_3 \nu    \right)
    \end{equation}
    The matrix valued one form $U'$ which satisfies
    \begin{equation}
        (\varphi \circ \dd)( v_0 N_0 + v_1 \partial_E N_0 + v_2 \partial_\mu N_0 +  v_3 \nu  ) = \dd \vec v + U' \vec v 
    \end{equation}
    is given by 
    \begin{equation}\label{prop::QGMC}
         U' =
        \begin{pmatrix}
            &  & & & 0 \\
            &  &U_E & & 0 \\
            &  & & & 0\\
            &0 &0 &0 &0 
        \end{pmatrix} \dd E 
        + 
        \begin{pmatrix}
            &  & & & 0 \\
            &  &U_\mu & & 0 \\
            &  & & & 0\\
            &0 &0 &0 &0 
        \end{pmatrix} \dd \mu 
    \end{equation}

\paragraph*{The quantum frame.}
The quantum action can be written as column vectors defined by the classical basis described above.
Using the differential operator $\mathcal{L}_{E,\mu}$ defined in equation \eqref{eq::DiffOp} the quantum action $ N$ decomposes according to $ N =   q_0 N_0 + q_1 \partial_E N_0 + q_2 \partial_\mu N_0 + \hbar \nu \mapsto \vec q $ where
\begin{equation}\label{eq::defqvec}
\begin{aligned}
    \vec q = &
    \begin{pmatrix}
        1 \\
        0 \\
        0\\
        0
    \end{pmatrix} +
    \begin{pmatrix}
        0 \\
        0 \\
        0\\
        1
    \end{pmatrix} \hbar
   + \Vast[ 
     \frac{1}{24} \begin{pmatrix}
        0\\
        5\\
        3\\
        0
        \end{pmatrix} -
\frac{1}{48(E-\mu)} \begin{pmatrix}
    1\\
    2\\
    2\\
    0
    \end{pmatrix} 
    \\ & \qquad \qquad \qquad -
    \frac{1}{3 \left(4 \mu+\left(E-1\right)^{2}\right)}\begin{pmatrix}
        \frac{1}{4}(E-1)\\
        \mu \\
        \frac{1}{2}(1-E)\mu\\
        0
        \end{pmatrix} 
     \Vast]\hbar^2 + \mathcal{O}(\hbar^4) .
\end{aligned}
\end{equation}

Let $\vec q$ be the vector defined in equation \eqref{eq::defqvec}.
Let $g$ be the matrix given by
    \begin{equation}\label{eq::gdef}\setlength\arraycolsep{6pt}
        g = \begin{pmatrix}
            \vec q & \partial_E \vec q + \hat U_E \vec q & \partial_\mu \vec q + \hat U_\mu \vec q & \partial_\hbar \vec q 
        \end{pmatrix}.
    \end{equation}
    The matrix $g$ in \eqref{eq::gdef} defines a change of frame from the classical basis 
    $$\{N_0, \partial_E N_0, \partial_\mu N_0 , \nu \}$$
    to the quantum basis 
     $$\{ N,\partial_E N ,\partial_\mu N ,\partial_\hbar N  \}.$$
    As a consequence this matrix $g$ has determinant given by the power series
    \begin{equation}\label{eq::detg}
        \det g = 1-\frac{1}{16} \frac{1}{\mu} \hbar^{2}+\frac{3}{512} \frac{1}{\mu^{2}} \hbar^{4}-\frac{5}{8192} \frac{1}{\mu^{3}} \hbar^{6}+\mathrm{O}\! \left(\hbar^{8}\right) 
    \end{equation}
    which can be seen by direct calculation.
    The matrix $g$ can be used to map the connection $U'$ defined by \eqref{prop::QGMC} to the corresponding connection $\hat U$ in the quantum frame (quantum basis).
    It is expressible as
    \begin{equation}
        \hat U = g^{-1} \dd g + g^{-1}  U' g
    \end{equation}
    according to the usual change of frame formula for a connection.
    The terms of the series appearing in equation \ref{eq::detg} can be seen to be generated by
    \begin{equation}
       \frac{1}{ \sqrt{1 + \frac{1}{8\mu} \hbar^2}}.
    \end{equation}

Let $N_{A},N_{B},N_{C}$ be the three independent quantum actions and choose $N_C$ to correspond to be the action associated to a pole of the potential. 
That is, $N_C = \sqrt{ 2\mu + \frac{1}{4}\hbar^2} + \nu \hbar $.
Let
\begin{equation}\setlength\arraycolsep{6pt}
    \hat M  =  \begin{pmatrix}
                 \hbar & 0 & 0 & 1  \\
                 N_{A} &  \partial_E N_{A}  &  \partial_\mu N_{A} & \partial_\hbar N_{A}\\
                 N_{B} &  \partial_E N_{B}  &  \partial_\mu N_{B} & \partial_\hbar N_{B}\\
                 N_{C} &  \partial_E N_{C}  &  \partial_\mu N_{C} & \partial_\hbar N_{C}\\
              \end{pmatrix}.
\end{equation}
Then the right hand side of the quantum bilinear identity \eqref{eq::SpheroidalBilinear} is given by
\begin{equation}
    \det \hat M \sqrt{ 2\mu + \frac{1}{4}\hbar^2}.
\end{equation}
It is easily verified that
    \begin{equation}
        \dd \hat M = \hat M \hat U.
    \end{equation}
    Therefore, applying Jacobi's formula for the derivative of a determinant then gives
    \begin{equation}
        \dd \det \hat M = \det \hat M \tr (\hat M ^{-1}\hat  M \hat U)= \tr \hat U \det \hat M.
    \end{equation}
    Let $c \in \C$.
    We have that
    \begin{equation}
        \begin{aligned}
            \det \hat M &= e^{ \int \tr \hat U}
             = e^{ \int \tr (g^{-1} \dd g + g^{-1}  U' g)}
             = e^{\tr \log g+  \int \tr U' }
             = \frac{ c \det g  }{\sqrt{2\mu}},
        \end{aligned}
    \end{equation}
    which means 
    \begin{equation}
        \det \hat M = \frac{c \det g}{\sqrt{2\mu}} .
    \end{equation}
    Combining this with the above observation that $\det g = (1 + \hbar^2/8\mu)^{-\frac{1}{2} } $ shows up to the desired order in $\hbar$ that
    \begin{equation}
        \det \hat M \sqrt{ 2\mu + \frac{1}{4}\hbar^2}  = c.
    \end{equation}
    which verifies the bilinear identity.

\subsection{From the bilinear relation to the P/NP relation}
\label{sec::FromBilinearToThePNPRelation}

Let $\tilde{N}_B(l,\mu,\hbar) = N_B(E_A(l,\mu,\hbar) , \mu ,\hbar)$ where $E_A(l,\mu,\hbar)$ comes from solving the Bohr-Sommerfeld condition such as in \eqref{eq::PerturbativeSeries}.
The above bilinear relation \eqref{eq::SpheroidalBilinear} is equivalent to the following P/NP relation
\begin{equation}\label{eq::PNPRelation}
    2 \frac{i}{ \hbar } \frac{\partial E_A}{\partial l}  = \tilde{N}_B -  2 \mu \frac{\partial \tilde{N}_B}{\partial \mu} - \hbar  \frac{\partial \tilde{N}_B}{\partial \hbar } .
\end{equation}
The bilinear identity \eqref{eq::SpheroidalBilinear} can be translated to a P/NP relation by evaluating it at $E=E_A (l,\mu,\hbar)$.
Note that
\begin{equation}
    \begin{aligned}
        &\frac{\partial N_A}{\partial E} \Big\lvert_{E=E_A} =  \frac{2 \pi \hbar}{\frac{\partial E_A}{\partial l} }, & &&\frac{\partial N_B}{\partial E} \Big\lvert_{E=E_A} &= \frac{\frac{\partial \tilde{N}_B}{\partial l}  }{\frac{\partial E_A}{\partial l} } ,
     \\ &\frac{\partial N_A}{\partial \mu} \Big\lvert_{E=E_A} = - 2\pi \hbar \frac{ \frac{\partial E_A}{\partial \mu}}{  \frac{\partial E_A}{\partial l}  }, & &&\frac{\partial N_B}{\partial \mu} \Big\lvert_{E=E_A} &=\frac{\partial \tilde{N}_B}{\partial \mu} -  \frac{\frac{\partial E_A }{\partial \mu}}{\frac{\partial E_A}{\partial l} }\frac{\partial \tilde{N}_B}{\partial l} ,
     \\ &\frac{\partial N_A}{\partial \hbar} \Big\lvert_{E=E_A} = 2\pi \left(l - \hbar \frac{ \frac{\partial E_A}{\partial \hbar}}{  \frac{\partial E_A}{\partial l}  } \right), & &&\frac{\partial N_B}{\partial \hbar} \Big\lvert_{E=E_A} &=\frac{\partial \tilde{N}_B}{\partial \hbar} -  \frac{\frac{\partial E_A }{\partial \hbar}}{\frac{\partial E_A}{\partial l} }\frac{\partial \tilde{N}_B}{\partial l} .
    \end{aligned}
\end{equation} 
Using these results to evaluate equation \eqref{eq::SpheroidalBilinear} at $E_A$ gives
\begin{equation}
    \begin{aligned}
        & -4 i \pi = 2\pi \hbar  \left(  l + 2 \mu  \frac{ \frac{\partial E_A}{\partial \mu}}{  \frac{\partial E_A}{\partial l}  } - \left(l - \hbar \frac{ \frac{\partial E_A}{\partial \hbar}}{  \frac{\partial E_A}{\partial l}  } \right) \right) \frac{\frac{\partial \tilde{N}_B}{\partial l}  }{\frac{\partial E_A}{\partial l} }
        \\ &- \left( \tilde N_B -2\mu \left(\frac{\partial \tilde{N}_B}{\partial \mu} -  \frac{\frac{\partial E_A }{\partial \mu}}{\frac{\partial E_A}{\partial l} }\frac{\partial \tilde{N}_B}{\partial l} \right) - \hbar \frac{\partial \tilde{N}_B}{\partial \hbar} -  \frac{\frac{\partial E_A }{\partial \hbar}}{\frac{\partial E_A}{\partial l} }\frac{\partial \tilde{N}_B}{\partial l}  \right)  \frac{2 \pi \hbar}{\frac{\partial E_A}{\partial l} }.
    \end{aligned}
\end{equation}
This simplifies to
\begin{equation}
    \begin{aligned}
         -2\frac{i}{ \hbar} \frac{\partial E_A}{\partial l} &=  \left(  2\mu \frac{\partial E_A}{\partial \mu}  + \hbar\frac{\partial E_A}{\partial \hbar}  \right)\frac{\frac{\partial \tilde{N}_B}{\partial l}  }{\frac{\partial E_A}{\partial l} } + \hbar \left(  \frac{\partial \tilde{N}_B}{\partial \hbar} -  \frac{\frac{\partial E_A }{\partial \hbar}}{\frac{\partial E_A}{\partial l} }\frac{\partial \tilde{N}_B}{\partial l}  \right) 
        \\ & -   \tilde{N}_B + 2\mu \left(\frac{\partial \tilde{N}_B}{\partial \mu} -  \frac{\frac{\partial E_A }{\partial \mu}}{\frac{\partial E_A}{\partial l} }\frac{\partial \tilde{N}_B}{\partial l} \right)  
        \\ & = - \tilde{N}_B +  2\mu \frac{\partial \tilde{N}_B}{\partial \mu} + \hbar  \frac{\partial \tilde{N}_B}{\partial \hbar } .
    \end{aligned}
\end{equation}
Which confirms equation \eqref{eq::PNPRelation}.
We remark that the P/NP relation \eqref{eq::PNPRelation} stated here is in agreement with the P/NP relation presented in \cite{Kemalresurgencedeformedgenus1curves}.
The work of \c{C}avu\c{s}o\u{g}lu et al \cite{Kemalresurgencedeformedgenus1curves} proposes a generalization of P/NP relations to genus one curves where the quantum action has poles with non-zero residues.
This class contains the spheroidal equation and as a result the P/NP relation presented in equation (3) of \cite{Kemalresurgencedeformedgenus1curves} can be seen to be equivalent to equation \eqref{eq::PNPRelation} with the appropriate change of notation.

Inspired by the analysis of both \'Alvarez and Casares in \cite{GabrielAlvarez2000} and Ba\c{s}ar et al in \cite{Basar_2017}, which provide P/NP relations for a cubic oscillator and the Mathieu equation respectively, we search for an equation which expresses the non-perturbative corrections to the energy level purely in terms of the perturbative series.
This is found by noting that the P/NP relation \eqref{eq::PNPRelation} is a partial differential equation which can be solved, for instance using a computer algebra system such as Mathematica.
Solving it produces the desired equation expressing $\tilde{N}_B$ in terms of $E_A$.
That is, up to an unkown function of the form $C(\hbar/ \sqrt{\mu},l)$ we have
\begin{equation}\label{eq::PNPRelationResult}
    \tilde{N}_B(l,\mu,\hbar) =  -\frac{i}{\hbar} \int \frac{\partial E_{A}}{\partial l}\left(l, \mu t , \hbar \sqrt{t } \right)\frac{\dd t}{t^2} + \sqrt{\mu}\, C \left(\frac{\hbar}{\sqrt{\mu}},l\right).
\end{equation}
To apply the P/NP relation \eqref{eq::PNPRelationResult} to calculations the ambiguity must be fixed.
This difficulty can be overcome by comparing to a direct calculation of $\tilde N_B$ term-by-term in $\hbar$.
The dual action $\tilde N_B$ can be calculated independently from the P/NP relation \eqref{eq::PNPRelation} by evaluating $(\mathcal{L}_{E, \mu} N_{0,B})$ at $E = E_A$.
This gives the following
\begin{equation}\label{eq::NBtilde}
   \begin{aligned}
    &\tilde N_B (E,\mu,l) 
    = i \sqrt{2} \left(2 \sqrt{\mu +1}-\sqrt{\mu } \left(2 \text{ arcsinh}\left(\sqrt{\mu }\right)+i \pi \right)\right)
   \\ & + i \hbar  \Bigg(- \left(l +\frac{1}{2} \right) + \left(l+\frac{1}{2}\right) \log \left(\frac{ \left(l+\frac{1}{2}\right) \hbar }{ 8\sqrt{2} (\mu +1)^{3/2}}\right)
     -\frac{1}{24 \left(l+\frac{1}{2}\right)} +  \frac{7}{2880 \left(l+\frac{1}{2}\right)^3} 
     \\ & -\frac{31 }{40320 \left(l+\frac{1}{2}\right)^5}+ \mathcal{O}\left(\frac{1}{l^7}\right)  \Bigg)
     - i\hbar ^2 \Bigg(\frac{ (14 \mu -3) \left(l+\frac{1}{2}\right)^2}{8 \sqrt{2} (\mu +1)^{3/2}} 
     \\ & - \frac{1}{96 \sqrt{2}} \left(-\frac{12 \left(2 \text{ arcsinh}\left(\sqrt{\mu }\right)- i\pi \right)}{\sqrt{\mu }}-\frac{34}{\sqrt{\mu +1}}+\frac{67}{(\mu +1)^{3/2}}\right) \Bigg)
     \\ & + i \hbar ^3 \left(\frac{5  (4 (\mu -5) \mu +1) \left(l+\frac{1}{2}\right)^3}{64 (\mu +1)^3}-\frac{ \left(4 \mu ^2+556 \mu -17\right) \left(l+\frac{1}{2}\right)}{256 (\mu +1)^3} \right)+\dots
   \end{aligned}
\end{equation}

A subtlety is that this expansion must be considered as an expansion in the double limit $l \to \infty$ and $\hbar \to 0$.
This is due to the singularities of the coefficient functions of the differential operator $\mathcal{L}_{E,\mu}$, in particular the poles at $E = \mu$.
However, careful treatement of the expansion shows that this only affects the linear term in $\hbar$.
This is visible in equation \eqref{eq::NBtilde} where all other terms are polynomials in $l$.
It is possible to conjecture the form of the term linear in $\hbar$ in equation \eqref{eq::NBtilde} which is a series in $(l+1/2)^{-1}$.
We propose, 
\begin{equation}\label{eq::NBtildeLinearTerm}
     [\hbar]\tilde N_B (E,\mu,l) = -i \log \left( \frac{\sqrt{\pi } 2^{\frac{7l}{2} + \frac{9}{4}} (\mu +1)^{\frac{3 l}{2}+\frac{3}{4}} }{\Gamma (l+1) \hbar ^{l+\frac{1}{2}}} \right) 
\end{equation}
which has been verified to terms of order $(l+1/2)^{-7}$.

Let $\mathcal{E}_{A}$ be $E_{A}$ with the terms constant, linear and quadratic in $\hbar$ removed.
Comparison with equations \eqref{eq::NBtilde} and \eqref{eq::NBtildeLinearTerm} suggest the correct form of the P/NP relation with the ambiguity fixed is given by
\begin{equation}\label{eq::PNPRelationFixed}
    \begin{aligned}
        &\tilde{N}_B(l,\mu, \hbar) = i 2 \sqrt{2} \left( \sqrt{\mu+1} -\sqrt{\mu} \text{ arcsinh}\left(\sqrt{\mu}\right) \right) -   \pi \sqrt{2 \mu+\frac{\hbar^2}{4}} 
        \\ & - i \hbar  \log \left( \frac{\sqrt{\pi } 2^{\frac{7l}{2} + \frac{9}{4}} (\mu +1)^{\frac{3 l}{2}+\frac{3}{4}} }{\Gamma (l+1) \hbar ^{l+\frac{1}{2}}} \right)
         - \frac{i}{ \hbar} \int_0^1 \frac{\partial \mathcal{E}_{A}}{\partial l}\left(l, \mu t , \hbar \sqrt{t } \right)\frac{\dd t}{t^2}.
    \end{aligned}
\end{equation}
It is necessary to remove the constant, linear and quadratic terms because, with the above bounds in the integral, they lead to divergences.

\section{Transseries and  Resurgent analysis}
\label{sec::Transseries}

Generically, power series fail to capture the full behavior of eigenvalues of Schr\"odinger style differential equations \cite{Dunne_2014}.
Perturbative series, such as the Rayleigh-Schr\"odinger series, are generically factorially divergent. 
The divergence of the series indicates the presence of non-perturbative contributions to the eigenvalues.
The eigenvalues of the spheroidal equation are given by a \textit{transseries} expansion of the form
\begin{equation}\label{eq::TransseriesForm}
    E_A(l, \mu, \hbar) = \sum_{r=0}^\infty c^{(0)}_r(l, \mu) \hbar^r  \pm  i \pi \sigma_l (\mu )  \hbar^{\beta} e^{- S(\mu)/\hbar }   \sum_{r= 0 }^{\infty} c^{(1)}_r(l, \mu) \hbar^r +\dots
\end{equation}
The function $\sigma_l (\mu )$ is the \textit{Stokes constant} and controls the structure of the transseries with variations of parameters \cite{Aniceto_2019}.
Here it is defined such that $c^{(1)}_0(l,\mu) =1$.
The $\pm i \pi $ appear due to the fact that the non-perturbative terms appear in the transseries as a result of Borel summation of the perturbative series.
In this perspective, the term is tied to a Borel singularity which lies on the real axis.

Resurgent analysis \cite{Costin_Book,Aniceto_2019} predicts that the large $k$ asymptotics of the $c^{(0)}_k(l,\mu)$ series coefficients encode the non-perturbative terms appearing in the transseries.
In particular, the coefficients $c^{(0)}_l$ have the following asymptotic expansion
\begin{equation}\label{eq::LargeOrderGrowth}
    c^{(0)}_r = \sum_{k=0}^{M-1} \sigma_l (\mu )  \frac{\Gamma(r- k - \beta)}{S^{r-k-\beta}} c^{(1)}_k + \mathcal{O}\left(\frac{\Gamma(r- M - \beta)}{S^{r-M-\beta}} \right)
\end{equation}
where $\beta \in \R$ and $S>0$.
This relation can be used to extract non-perturbative information from the perturbative expansion numerically.
By rearranging terms in \eqref{eq::LargeOrderGrowth} it is possible to isolate $c^{(1)}_j$.
That is,
\begin{equation}\label{eq::IterativeFormula}
    c^{(1)}_j =  \lim_{r \to \infty} \frac{S^{r-j-\beta}}{ \sigma_l (\mu ) \Gamma(r- j - \beta)} \left( c^{(0)}_r - \sum_{k=0}^{j-1} \frac{\Gamma(r- k - \beta)}{S^{r-k-\beta}} c^{(1)}_k \right) .
\end{equation}
Relations of this form are described in \cite{Costin_Book} as well as in \cite{Aniceto_2019}. 
Given a large number of the perturbative coefficients $c^{(0)}_j$, as well as the first $(j-1)$ fluctuation coefficients $c^{(1)}_j$, it is possible to numerically approximate the limit.
This leads to a recursive algorithm for computing the fluctuation coefficients.
The numeric approximation of the limit can be improved using Richardson acceleration, see for example \cite{alma992542403502432}.
Richardson acceleration was used extensively in calculations described in section \ref{sec::NumericalTests} to develop a numeric approximation for the asymptotics of the spheroidal eigenvalues.

\subsection{The Transseries expansion of the spheroidal eigenvalue}
\label{sec::TransseriesExpansion}

The exact WKB method suggests that the correction to the perturbative energy level is accessible from a correction to the Bohr-Sommerfeld conditon \eqref{eq::BohrSommerfeld} of the form
\begin{equation}\label{eq::CorrectedBScondition}
    2\pi \hbar l =  N_A + \hbar\, q(N_B)
\end{equation}
where $q$ is exponentially small \cite{GabrielAlvarez2000,Ito_2019}.
Moreover, to first exponential order we can expect that $q \sim  \pm e^{\frac{i}{\hbar} N_B}$, see for instance the discussion of the quartic oscillator in \cite{Ito_2019}.
By considering $q$ as a small correction to the level number $l$, we find the energy level with the first exponential correction included to be of the form
\begin{equation}\label{eq::FormOfNPcorrections}
    E_A(l,\mu,\hbar) = \sum_{r=0}^\infty c^{(0)}_r(l, \mu) \hbar^r \mp  \frac{i}{2 \pi } \pdv{E_A}{l}  e^{\frac{i}{\hbar} \tilde N_B}  +\dots 
\end{equation}
An important remark is that the precise meaning of this non-perturbative correction is set by boundary conditions.
See for instance \cite{Misumi_2025,Imaizumi::2020} which addresses the exact quantization condition of the Mathieu equation with periodic boundary conditions. 
In this context, the two cases would identify the top and bottom of a band of the spectrum \cite{Basar_2017,Misumi_2025}.
With this we can write the fluctuation series using the P/NP relation \eqref{eq::PNPRelationFixed}.
It is given by the following,
\begin{equation}\label{eq::FluctuationSeries}
      \sum_{r=0}^\infty c^{(1)}_r(l,\mu) \hbar^r 
        = \frac{1}{\hbar \sqrt{2(\mu + 1 )}} \frac{\partial E_A}{\partial l} e^{ \frac{i}{\hbar} \left( -\frac{i}{\hbar} \int_0^1 \frac{\partial \mathcal{E}_{A}}{\partial l}\left(l, \mu t , \hbar \sqrt{t } \right)\frac{\dd t}{t^2}  +  \pi \sqrt{2 \mu +\frac{\hbar^2}{4}} -  \pi \sqrt{2 \mu }  \right)} .
\end{equation}
These coefficients were calculated up to $r=4$ using equation \eqref{eq::FluctuationSeries}.
The coefficients $c_0^{(1)},c_1^{(1)}$ and $c_2^{(1)}$ are listed in appendix \ref{sec::FluctuationCoefficients}.
Further justification for equation \eqref{eq::FluctuationSeries} and the above discussion come from the analysis of the P/NP relation of the Mathieu equation which is described in \cite{Basar_2017}.
The limit $\mu \to 0$ of the spheroidal equation reproduces the Mathieu equation and, as is discussed in section \ref{sec::LimitToMathieu}, the limit of equation \eqref{eq::FluctuationSeries} reproduces the corresponding result for the Mathieu equation.

The remaining parts of the transseries were found to be the following:
The Stokes constant $\sigma_l( \mu )$ introduced in equation \eqref{eq::TransseriesForm} is determined by the expression for $[\hbar^1]\tilde N_B$ given in equation \eqref{eq::NBtildeLinearTerm} and the density of states $\partial E_A/ \partial l$ to be the following
\begin{equation}\label{eq::StokesConstantFormula}
    \sigma_l (\mu ) = -\frac{2^{\frac{7 }{2} l +\frac{7}{4}} (\mu+1)^{\frac{3 l}{2}+\frac{5}{4}}}{\pi ^{3/2} \Gamma (l+1)}.
\end{equation}
We emphasize that this result is supported by the numerical experiments described in section \ref{sec::NumericalTests}.
The parameter $\beta$ in \eqref{eq::TransseriesForm} is given according to equation \eqref{eq::FormOfNPcorrections} by
\begin{equation}
    \beta = \frac{1}{2}-l
\end{equation}
The function $S(\mu)$ in \eqref{eq::TransseriesForm} is given by the classical dual action evaluated at the bottom of the well $E=\mu$.
That is
\begin{equation}\label{eq::Saction}
    S(\mu) =  2\sqrt{2}  \left(  \sqrt{\mu+1}- \sqrt{\mu} \text{ arcsinh}\left(\sqrt{\mu}\right)\right) + i\pi \sqrt{2 \mu}.
\end{equation}

\subsection{Limit to the Mathieu Equation}
\label{sec::LimitToMathieu}
Taking the limit  $\mu \to 0$ reduces the transseries expansion \eqref{eq::TransseriesForm} to that of the Mathieu spectrum.
However, due to the poles of the spheroidal potential taking the limit is non-trivial.
In particular, the series expansion 
\begin{equation}\label{eq::QuantumResidue}
    \sqrt{ 2\mu + \frac{1}{4}\hbar^2} =\sqrt{2} \sqrt{\mu }+\frac{\hbar ^2}{8 \sqrt{2} \sqrt{\mu }} - \frac{\hbar ^4}{256 \sqrt{2} \mu ^{3/2}} + \frac{\hbar ^6}{4096 \sqrt{2} \mu ^{5/2}}+O\left(\hbar ^7\right)
\end{equation}
which describes the residue associated to a pole, produces divergences in the limit $\mu \to 0$.
By summing this series the limit can be taken in a well-defined manner.
With this observation the limit to the transseries of the Mathieu equation can be completed by first subtracting the `divergent' terms from the quantum dual action. 
The correct limit can be found from 
\begin{equation}
    \begin{aligned}
    & \lim_{\mu \to 0} \frac{\partial E_A}{\partial l} e^{ \frac{i}{\hbar}\tilde{N}_B } = e^{ \frac{i \pi}{2} } \lim_{\mu \to 0}  \frac{\partial E_A}{\partial l} e^{ \frac{i}{\hbar}\left(\tilde{N}_B -   \pi \sqrt{ 2\mu + \frac{1}{4}\hbar^2} \right)} 
    \\ & = - i \frac{ 2^{\frac{7 l}{2}+\frac{7}{4}} \hbar^{l-\frac{1}{2}}  e^{-\frac{2\sqrt{2}}{\hbar}} }{\sqrt{\pi } \Gamma (l+1)}
      \Bigg(1 + \frac{\left(-4 \left(l+\frac{1}{2}\right) \left(3 \left(l+\frac{1}{2}\right)+4\right)-3\right) \hbar }{32 \sqrt{2}}
    \\ &+\frac{\left(8 \left(l+\frac{1}{2}\right) \left(\left(l+\frac{1}{2}\right) \left(2 \left(l+\frac{1}{2}\right) \left(9 \left(l+\frac{1}{2}\right)+4\right)-39\right)-22\right)-87\right) \hbar ^2}{4096} +\dots
    \Bigg)
\end{aligned}
\end{equation}
which agrees with the non-perturbative structure coming from the Mathieu equation listed in \cite{Basar:2015xna}.

\subsection{Numerical tests}
\label{sec::NumericalTests}

The relation between transseries expansions and asymptotics given by equation \eqref{eq::LargeOrderGrowth} provides a numerical check on the form of the transseries expansion given in equation \eqref{eq::TransseriesForm}.
The tests utilizes the BenderWu mathematica package \cite{Sulejmanpasic_Unsal_2016} to calculate the perturbative expansion of the energy level to high orders with $\mu,l$ set to fixed, numeric values.
Using this data, Richardson extrapolation combined with equation \eqref{eq::IterativeFormula} can be used to extract both the Stokes constant and the non-perturbative series $c_k^{(1)}(l,\mu)$ coefficients to high precision.

\begin{figure}
    \includegraphics[width = 14cm]{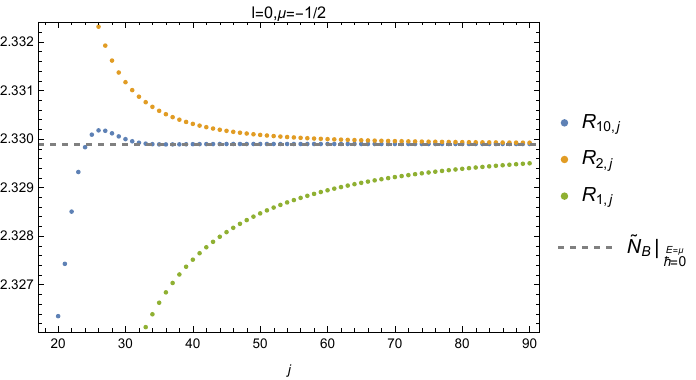}
    \caption{\label{fig::RichardsonConvergence}
    The plot shows the convergence acceleration provided by Richardson extrapolation.
    The sequences $R_{i,j}$ are the $i$th order Richardson of the ratio $r_j$ defined by \eqref{eq::Ratios}.
    The gray dashed line is $ N_B \lvert_{\hbar=0,E=\mu}$ with $\mu = -1/2$. }
\end{figure}

\begin{figure}
    \includegraphics[width = 14cm]{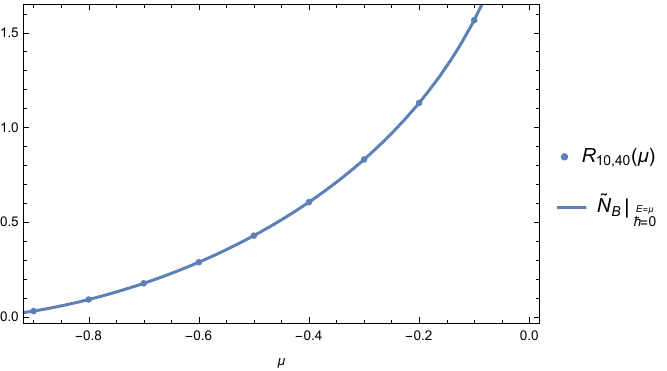}
    \caption{\label{fig::ActionFigure}
    The sequence $R_{10,40}$ is the $10$th order Richardson extrapolation of the ratio $r_j$ defined by \eqref{eq::Ratios} evaluated at $j=40$.
    The solid blue curve is $\tilde N_B\lvert_{\hbar=0,E=\mu} =  S(\mu)$ given by equation \eqref{eq::Saction}. }
\end{figure}

\begin{figure}
    \includegraphics[width = 14cm]{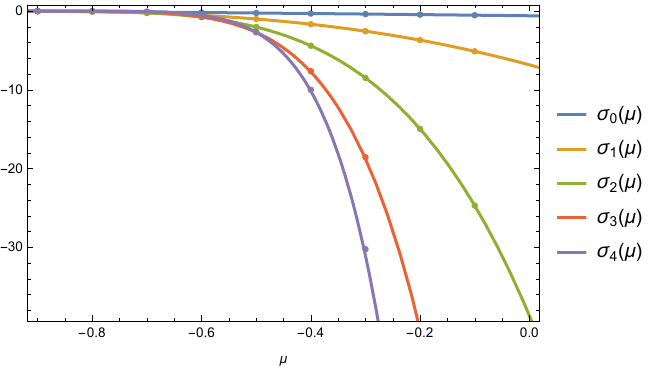}
    \caption{\label{fig::ScFigure} The plot shows the numerically calculated and exact Stokes constant given by $\sigma_l(\mu)$ defined in equation \eqref{eq::StokesConstantFormula}. 
    The dots are the numerically calculated values while the solid lines are the exact values.
    The blue, orange, green, red and purple correspond to $l=0,1,2,3,4$ respectively. 
    Each numeric approximation used 50 perturbative coefficients and Richardson acceleration of order 10.
    }
\end{figure}

According to equation \eqref{eq::LargeOrderGrowth} the ratios 
\begin{equation}\label{eq::Ratios}
   \begin{aligned}
    &r_j = \frac{c^{(0)}_j(l,\mu)}{j c^{(0)}_{j-1}(l,\mu)}, & \text{ and }  &&\delta_j = \left(S(\mu)\right)^{l-\frac{1}{2}+j} \frac{c^{(0)}_j(l,\mu)}{\Gamma(j+ l -1/2)}
   \end{aligned}
\end{equation}
converge to the action $S(\mu)^{-1}$ and the Stokes constant $\sigma_l(\mu)$, respectively.
By accelerating the convergence with Richardson extrapolation the numerically calculated action was found to agree precisely with the exact calculation.
This is illustrated in figure \ref{fig::ActionFigure} which show the agreement between the numerically calculated values.
Figure \ref{fig::RichardsonConvergence} shows the convergence of $r_j$ and its Richardson extrapolation to the action for $\mu =-1/2$.
The agreement of the numeric prediction of the Stokes constant is illustrated in figure \ref{fig::ScFigure} which shows the numerically calculated values of $\sigma_l$ alongside the exact value described by equation \eqref{eq::StokesConstantFormula}.

\begin{figure}
    \includegraphics[width = 14cm]{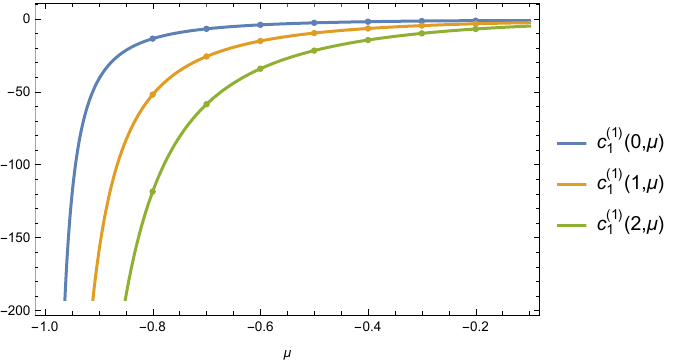}
    \caption{\label{fig::c1Figure} The plot shows the numerical and exact fluctuation coefficient $c^{(1)}_1(l,\mu)$. 
    The dots are the numerically calculated values while the solid lines are the exact prediction.
    The blue, orange and green correspond to $l=0,1,2$ respectively. 
    Each numeric approximation used 100 series coefficients and Richardson acceleration of order 10.
    }
\end{figure}

\begin{figure}
    \includegraphics[width = 14cm]{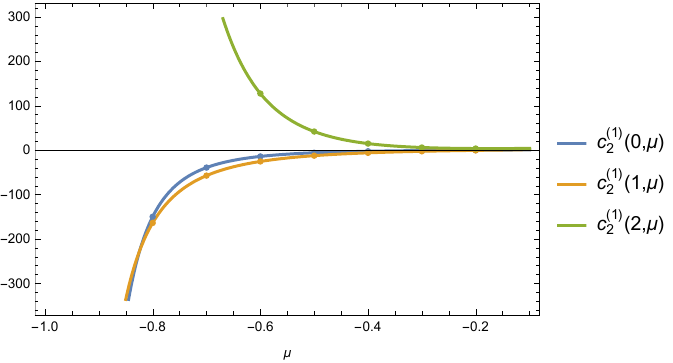}
    \caption{\label{fig::c2Figure} The plot shows the numerical and exact fluctuation coefficient $c^{(1)}_2(l,\mu)$. 
    The dots are the numerically calculated values while the solid lines are the exact prediction.
    The blue, orange, and green correspond to $l=0,1,2$ respectively. 
    Each numeric approximation used 100 series coefficients and Richardson acceleration of order 10.
    }
\end{figure}

\begin{figure}
    \includegraphics[width = 14cm]{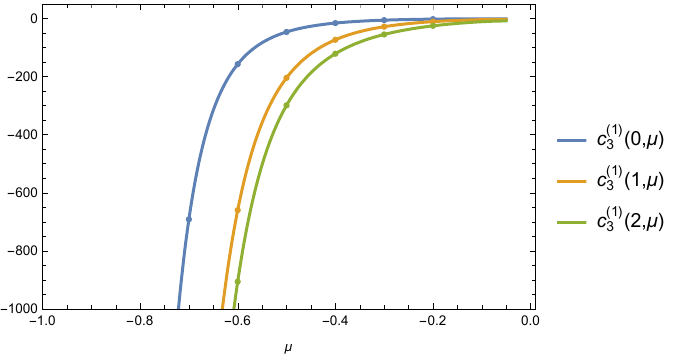}
    \caption{\label{fig::c3Figure} The plot shows the numerical and exact fluctuation coefficient $c^{(1)}_3(l,\mu)$. 
    The dots are the numerically calculated values while the solid lines are the exact prediction.
    The blue, orange, and green correspond to $l=0,1,2$ respectively.
    Each numeric approximation used 100 series coefficients and Richardson acceleration of order 10.
    }
\end{figure}

\begin{figure}
    \includegraphics[width = 14cm]{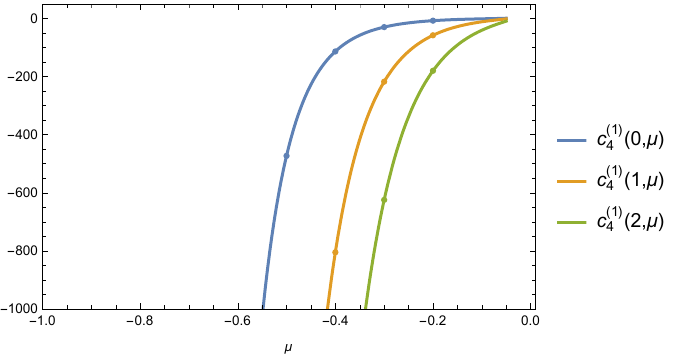}
    \caption{\label{fig::c4Figure}  The plot shows the numerical and exact fluctuation coefficient $c^{(1)}_4(l,\mu)$. 
    The dots are the numerically calculated values while the solid lines are the exact prediction.
    The blue, orange, and green correspond to $l=0,1,2$ respectively.
    Each numeric approximation used 100 series coefficients and Richardson acceleration of order 10.
    }
\end{figure}

In addition, the fluctuation coefficients were calculated by applying equation \eqref{eq::IterativeFormula} iteratively.
These numeric calculations were done for $j=1,2,3,4$ and are illustrated in figures \ref{fig::c1Figure}, \ref{fig::c2Figure}, \ref{fig::c3Figure} and  \ref{fig::c4Figure}.
The agreement between the numeric calculations and the exact fluctuation series \eqref{eq::FluctuationSeries} confirms the proposed non-perturbative structure.


\section{Conclusion}
The analysis presented above provides new non-perturbative results for the spheroidal eigenvalues as well as a parametric-resurgent analysis of the perturbative spectral data.
We provided an algorithmic approach to the calculation of the quantum actions as well as a novel analysis of the P/NP relation that they satisfy.
Finally, we presented a numerical Resurgent analysis which verified our results.

There are several directions for future research: 
(1) Analyze the interplay between the perturbative and non-perturbative structures of the eigenvalues for various regions of parameters, including inverting the potential, $\hbar\mapsto i \hbar$, as well as rotating to the imaginary axis, $x\mapsto ix$.
(2) Developing exact quantization conditions and solving them with both TBA methods as well as transseries. 
(3) Generalize the analysis to treat the spin weighted spheroidal equation.
(4) Study the non-perturbative structures in the quasinormal modes of Kerr and Kerr-Newman black holes where the spheroidal eigenvalues play a fundamental role \cite{Aminov2020,TeukolskyThesis,Silva_2025}.

\section{Acknowledgements}
This work is supported in part by the U.S. Department of Energy, Office of High Energy Physics, Award DE-SC0010339.
The author would like to thank Gerald Dunne and Hal Haggard for many insightful comments and valuable feedback, as well as Kemal Tezgin for correspondence.

\appendix

\section{The complete Elliptic integrals}
\label{sec::EllipticIntegrals}

The complete elliptic integrals $\mathbf{K},\mathbf{E}$ and $\mathbf{\Pi}$ are defined as follows
\begin{equation}
    \begin{aligned}
        &\mathbf{K}(k) = \int_0^1 \frac{dz}{\sqrt{(1-z^2)(1- k^2 z^2)}}, 
        \\ &\mathbf{E}(k) = \int_0^1 \sqrt{\frac{1- k^2 z^2}{1-z^2}}dz, 
        \\ &\mathbf{\Pi}(\nu,k) = \int_0^1 \frac{dz}{(1-\nu z^2)\sqrt{(1-z^2)(1- k^2 z^2)}}.
    \end{aligned}
\end{equation}
They satsify the following differential equations \cite{NIST:DLMF}:
\begin{equation}
     \begin{aligned}
          & \left(k^{2}-1\right) \frac{\partial^{2} \mathbf{\Pi} }{\partial k^{2}} + \frac{\left(3 k^{4}- \left(\nu  + 1\right) k^{2}-\nu \right)}{k \left(k^{2}-\nu \right)} \frac{\partial \mathbf{\Pi}}{\partial k}  -\frac{2 \left(-1+\nu \right) \nu }{k^{2}-\nu}\frac{\partial\mathbf{\Pi}}{\partial \nu} +\mathbf{\Pi}=0,
       \\ & \frac{\left(k^{2}-\nu\right) }{k} \frac{\partial^{2} \mathbf{\Pi}}{\partial \nu \partial k} -\frac{1}{k}\frac{ \partial^{2} \mathbf{\Pi}}{\partial \nu^{2}}  +\frac{\partial \mathbf{\Pi}}{\partial \nu} =0,
       \\ & \left(k^{2}-1\right) \frac{d^{2}\mathbf{K} }{d k^{2}} +\frac{\left(3 k^{2}-1\right)}{k} \frac{d \mathbf{K}}{d k} + \mathbf{K}=0,
       \\ & \left(k^{2}-1\right) \frac{d^{2}\mathbf{E} }{d k^{2}} +\frac{\left(1- k^{2}\right)}{k} \frac{d \mathbf{E}}{d k} + \mathbf{E}=0.
    \end{aligned}
\end{equation}

\section{The Differential Operator}
\label{sec::LEmu}

The differential operator of equation \eqref{eq::DiffOp} can be written
\begin{equation}
    \mathcal{L}_{E,\mu} = q_0(E,\mu) + q_E(E,\mu) \pdv{}{E} + q_\mu (E,\mu) \pdv{}{\mu}
\end{equation}
where $q_0 ,q_E$ and $q_\mu$ are given by up to order $\hbar^8$ by the following three expressions:
\tiny
\begin{equation}
    \begin{aligned}
        & q_0(E,\mu) = 1 +\hbar ^2 \left(\frac{1-E}{12 (4\mu + (E-1)^2 )}+\frac{1}{48 \mu-48 E}\right) 
        \\ &+\hbar ^4 \Bigg(\frac{-15 E^2+92 E-33}{7680  \left(E+1\right)^3 (\mu-E)}
        +\frac{101 E^3+57 E^2-113 E-45}{960( E+1) \left(4 \mu +(E-1)^2\right)^2}
        +\frac{-7 E^4+14 E^3-14 E+7}{90 \left(4 \mu +(E-1)^2\right)^3}
        \\ & +\frac{-25 E^4-50 E^3-76 E^2+90 E+1}{960 (E-1) (E+1)^3 \left(4 \mu +(E-1)^2\right)}
         +\frac{-7 E-7}{2880 (\mu-E)^3}
        +\frac{-34 E-11}{7680( E+1) (\mu-E)^2}
        +\frac{1}{512 \mu ( E-1)}\Bigg)
        \\ & + \hbar ^6 \Bigg(-\frac{1}{8192 \mu^2 ( E-1)}
        +\frac{5-E}{24576 \mu (E-1)^3}
        +\frac{31 E^2+62 E+31}{20160 (\mu-E)^5}
         +\frac{3256 E^2-3000 E+3933}{2580480\left( E+1\right)^2 (\mu-E)^3}
       \\ &  +\frac{210 E^3-4715 E^2+7424 E-11327}{2580480 \left(E+1\right)^4 (\mu-E)^2}
         +\frac{35 E^4+5964 E^3-36066 E^2+21764 E+10559}{860160 (E+1)^6 (\mu-E)}
        \\ & +\frac{4425 E^5-8549 E^4+1042 E^3+3918 E^2+1109 E-1945}{5040 \left(4 \mu +(E-1)^2\right)^4}
        \\ & +\frac{-97917 E^5-95727 E^4+85758 E^3+110082 E^2-11089 E+8893}{161280 (E+1)^2 \left(4 \mu +(E-1)^2\right)^3}
        \\ & +\frac{19773 E^6+53598 E^5+18015 E^4-13648 E^3-102181 E^2+25306 E+2497}{161280 (E-1) (E+1)^4 \left(4 \mu +(E-1)^2\right)^2}
        \\ & +\frac{-364 E^6+3333 E^5-9359 E^4+10370 E^3-4506 E^2-683 E+649}{13440 (E-1)^3 (E+1)^6 \left(4 \mu +(E-1)^2\right)}
        \\ & +\frac{-124 E^7+372 E^6-124 E^5-620 E^4+620 E^3+124 E^2-372 E+124}{315 \left(4 \mu +(E-1)^2\right)^5}
         +\frac{914 E+1031}{322560 (\mu-E)^4}\Bigg)
        ,
    \end{aligned}
\end{equation}

\begin{equation}
    \begin{aligned}
      \\ &q_E(E,\mu) =  \hbar ^2 \left(\frac{(E-1)^2}{12 \left(4 \mu+(E-1)^2\right)}-\frac{E}{24 (\mu-E)}+\frac{1}{12}\right)
       +\hbar ^4 \Bigg(\frac{102 E^2+61 E+28}{11520( E+1) (\mu-E)^2}
      \\ & +\frac{E^4+18 E^2-8 E+1}{192 (E+1)^3 \left(4 \mu + (E-1)^2\right)}
       +\frac{3 E^3-15 E^2+9 E-1}{768  (E+1)^3 (\mu-E)}
      +\frac{-247 E^4+20 E^3+510 E^2-92 E-191}{2880( E+1) \left(4 \mu + (E-1)^2\right)^2}
      \\ & +\frac{7 E^5-21 E^4+14 E^3+14 E^2-21 E+7}{90 \left(4 \mu + (E-1)^2\right)^3}+\frac{7 E (E+1)}{1440 (\mu-E)^3}-\frac{1}{256 \mu}\Bigg)
       + \hbar^6 \Bigg(\frac{1}{4096 \mu^2}
     \\ & +\frac{E-3}{\mu 12288 \left( E-1\right)^2}
      +\frac{-914 E^2-1279 E-248}{161280 (\mu-E)^4}
       -\frac{31 E (E+1)^2}{10080 (\mu-E)^5}
     \\ &  +\frac{-3256 E^3-656 E^2-9481 E-1892}{1290240 \left( E+1\right)^2 (\mu-E)^3}
      +\frac{-210 E^4+3087 E^3-3439 E^2+14777 E-2163}{1290240 (E+1)^4 (\mu-E)^2}
     \\ & +\frac{-105 E^5-17997 E^4+112282 E^3-75414 E^2-35313 E+10675}{1290240 (E+1)^6 (\mu-E)}
     \\ & +\frac{66981 E^6+12946 E^5-99109 E^4-64004 E^3+58635 E^2+4562 E+19989}{\left( E+1\right)^2 \left(4 \mu + (E-1)^2\right)^3}
     \\ & +\frac{-1416 E^6-2790 E^5+6729 E^4-23236 E^3+27016 E^2-6602 E-1381}{80640 (E+1)^4 \left(4 \mu + (E-1)^2\right)^2}
     \\ & +\frac{-3929 E^6+10990 E^5-7111 E^4-2876 E^3+329 E^2+5038 E-2441}{5040 \left(4 \mu + (E-1)^2\right)^4}
     \\ & +\frac{124 E^8-496 E^7+496 E^6+496 E^5-1240 E^4+496 E^3+496 E^2-496 E+124}{315 \left(4 \mu + (E-1)^2\right)^5}
     \\ & +\frac{-1845 E^8-7380 E^7+1356 E^6-66548 E^5+178750 E^4-104156 E^3-8276 E^2+21844 E-7025}{161280 \left(E-1\right) (E+1)^6 \left(4 \mu + (E-1)^2\right)}\Bigg)  
     ,
    \end{aligned}
\end{equation}

\begin{equation}
    \begin{aligned}
        \\ &q_m(E,\mu) =  \hbar ^2 \left(\frac{E}{24}+\frac{1}{24}-\frac{(E-1)^3}{24 \left(4 \mu+(E-1)^2\right)}-\frac{E}{24 (\mu-E)}\right)
        + \hbar ^4 \Bigg(\frac{158 E^2+145 E+56}{(11520 E+11520) (\mu-E)^2}
        \\ & +\frac{49 E^3-13 E^2+89 E+11}{3840(E+1)^3 (\mu-E)}
        +\frac{-7 E^6+28 E^5-35 E^4+35 E^2-28 E+7}{180 \left(4 \mu + (E-1)^2 \right)^3}
        +\frac{7 E (E+1)}{1440 (\mu-E)^3}-\frac{1}{256 \mu}
         +\frac{5}{384}
        \\ & +\frac{527 E^5-659 E^4-826 E^3+1162 E^2+155 E-359}{5760( E+1) \left(4 \mu + (E-1)^2 \right)^2}
         +\frac{-63 E^5-142 E^4-64 E^3+190 E^2+57 E+22}{960 (E+1)^3 \left(4 \mu + (E-1)^2 \right)}
         \Bigg)
        \\ &  + \frac{\hbar ^6}{1290240} \Bigg(\frac{315}{\mu^2}
         +\frac{128 \left(6409 E^2+8694 E+3929\right) (E-1)^5}{\left(4 \mu+(E-1)^2\right)^4}
         -\frac{8 \left(1410 E^2+2023 E+496\right)}{(\mu-E)^4}
       \\ & +\frac{10568 E^3+19872 E^2+27741 E+8248}{(E+1)^2 (E-\mu)^3}
        -\frac{3968 E (E+1)^2}{(\mu-E)^5}-\frac{315}{\mu-\mu E}
        -\frac{253952 (E+1)^2 (E-1)^7}{\left(4 \mu+(E-1)^2\right)^5}
        \\ &  -\frac{4 \left(239517 E^4+628076 E^3+614190 E^2+273516 E+71133\right) (E-1)^3}{(E+1)^2 \left(4 \mu+(E-1)^2\right)^3}
        \\ & +\frac{-3466 E^4+1203 E^3-8613 E^2+6461 E-3933}{(E+1)^4 (\mu-E)^2}
        \\ & +\frac{-315 E^5-13597 E^4+109994 E^3-64098 E^2-16447 E+11327}{(E+1)^6 (\mu-E)}
        \\ &  +\frac{8 \left(58845 E^7+162693 E^6+84015 E^5-108767 E^4-191683 E^3+15345 E^2-14753 E-5695\right)}{(E+1)^4 \left(4 \mu+(E-1)^2\right)^2}
        \\ &  -\frac{4 \left(19773 E^8+93144 E^7+149352 E^6+35984 E^5+846 E^4-317144 E^3+5000 E^2+38496 E-5291\right)}{(E-1) (E+1)^6 \left(4 \mu+(E-1)^2\right)}
        \Bigg)
          ,
    \end{aligned}
\end{equation}

\normalsize
\section{The Fluctuation Coefficients}
\label{sec::FluctuationCoefficients}

The first three of the calculated fluctuation coefficients defined in \eqref{eq::FluctuationSeries} are listed below
\begin{equation}
   \begin{aligned}
    c^{(1)}_0 &=1,
\\   c^{(1)}_1 &= \frac{7 \mu  \left(l+\frac{1}{2}\right)^2}{4 \sqrt{2} (\mu +1)^{3/2}}
    -\frac{3 \left(l+\frac{1}{2}\right)^2}{8 \sqrt{2} (\mu +1)^{3/2}}
    +\frac{\mu  \left(l+\frac{1}{2}\right)}{\sqrt{2} (\mu +1)^{3/2}}
    -\frac{l+\frac{1}{2}}{2 \sqrt{2} (\mu +1)^{3/2}}
   \\ & -\frac{11}{32 \sqrt{2} (\mu +1)^{3/2}}
    +\frac{17 \mu }{48 \sqrt{2} (\mu +1)^{3/2}}
    -\frac{i \pi }{8 \sqrt{2} \sqrt{\mu }}+\frac{\text{arcsinh}\left(\sqrt{\mu }\right)}{4 \sqrt{2} \sqrt{\mu }} ,
\\ c^{(1)}_2 &= \frac{1}{18432 \sqrt{2} \mu (\mu+1)^{5/2}} \Bigg( \Big(  24 i \pi \sqrt{\mu}  \sqrt{\mu+1} (33  -   \mu - 34  \mu^{2})  -144 \pi ^2 (\mu+1)^3
\\ &+ (4 \mu (865 \mu+4623)+225) \mu 
\Big)  
  + 48 \sqrt{\mu}\Big( \sqrt{\mu} (4 \mu (37 \mu+367)+15) 
 \\ &-24 i \pi  \sqrt{\mu+1} \left(2 \mu^2+\mu-1\right) \Big) \left(l +\frac{1}{2}\right)
 + 24 \sqrt{\mu } \Big( (4 \mu  (119 \mu +435)-45) \sqrt{\mu } 
 \\ & -12 i \pi  (\mu +1)^{3/2} (14 \mu -3) \Big)\left(l +\frac{1}{2}\right)^2
 + 576 \mu  (12 \mu  (3 \mu +5)+1)\left(l +\frac{1}{2}\right)^3 
 \\ & +144 (3-14 \mu )^2 \mu\left(l +\frac{1}{2}\right)^4
 \Bigg) + 
 \Bigg( \frac{1}{384 \sqrt{2} \mu } \left(\frac{\sqrt{\mu } (34 \mu -33)}{\mu +1}-12 i \pi  \sqrt{\mu +1}\right)
 \\ & + \frac{2 \mu -1}{8 \sqrt{2} \sqrt{\mu } (\mu +1)}\left(l +\frac{1}{2}\right) 
 +\frac{14 \mu -3}{32 \sqrt{2} \sqrt{\mu } (\mu +1)} \left(l +\frac{1}{2}\right)^2    \Bigg) \text{ arcsinh}(\sqrt{\mu})
 \\ & +\frac{\sqrt{\mu+1} \text{ arcsinh}\left(\sqrt{\mu}\right)^2}{32 \sqrt{2} \mu}
   \end{aligned}
\end{equation}

\bibliography{refs.bib}

\end{document}